\documentclass[a4paper]{jpconf}

\usepackage{textcomp}
\usepackage{lmodern}
\usepackage{amsmath}
\usepackage{amssymb}
\usepackage{graphicx}
\usepackage{nomencl}
\usepackage{cite}
\usepackage{float}
\usepackage[pdftex]{hyperref}

\PassOptionsToPackage{hyphens}{url}\usepackage{hyperref}

\begin{document}

\title{Symmetry Breaking in UCA-Based Vortex Waves}
\author{Andrea Cagliero$^1$}
\address{$^1$ Microwaves Department of IMT Atlantique, Institut Mines-T\'{e}l\'{e}com; Lab-STICC (CNRS), Laboratory for Science and Technologies of Information, Communication and Knowledge, Brest, F-29238, France}
\ead{andrea.cagliero@edu.unito.it}

\address{\textbf{\\The present manuscript has been accepted for publication in ``Journal of Physics Communications'', DOI: 10.1088/2399-6528/aade74. It is available for reuse under a CC BY-NC-ND 3.0 licence after the 12 month embargo period provided that all the terms of the licence are adhered to.}}

\begin{abstract}
In literature, it is often said that a uniform circular array (UCA)
composed of a sufficient number of radiating elements is able to generate
a vortex wave with arbitrary topological charge. While providing a
very simple and intuitive solution without the need for any synthesis
method, the excitation of UCAs by means of a progressive phase shift
does not always guarantee that vortex waves are correctly generated,
since a major role is played by the symmetry and radiation properties
of each individual element which constitutes the array. When the UCAs
are employed in vortex communication links, these symmetry properties
strongly affect the orthogonality of the channel modes and, therefore,
the communication efficiency. In this work, the symmetry breaking
associated with the radiation of vortex waves by UCAs and its impact
on the mode orthogonality are explored in detail. 
\end{abstract}

\section{Introduction}

In the last years, electromagnetic waves characterized by a vortex
term $\varPhi_{m}\left(\varphi\right)=\exp\left(im\varphi\right)$,
where $\varphi$ is the azimuthal angle and the integer $m$ represents
the so-called topological charge, have become extremely popular, especially
by virtue of their orbital angular momentum (OAM) content (for a review
of the applications, see \cite{Allen2003,Andrews2008,Andrews2013,Torres2011}
and references therein). Vortex waves are usually described as solutions
of the homogeneous Helmholtz equation, often in the paraxial regime.
Since all these source-free modes just correspond to mathematical
idealizations, the problem of generating realistic vortex waves has
been largely addressed at both optical and radio frequencies.

In optics, vortex waves are synthesized in a variety of possible ways,
ranging from computer generated holograms \cite{Bazhenov1990} to
cylindrical lenses \cite{Beijersbergen1993} and spiral phase plates
\cite{Beijersbergen1994}. Most of the techniques involve the transformation
of conventional laser radiation in waves carrying OAM by means of
suitable optical devices and mode converters, the former being able
to imprint the characteristic phase topology on the incoming standard
beam through an azimuthal variation of the refractive index, the latter
exploiting the completeness of the modes (see also \cite{Padgett2000,Yao2011}).
Recently, many researchers have also dealt with the conversion of
the electromagnetic angular momentum from spin to orbital, bringing
to light the concept of \textit{q-plate} as a further technique for
the generation of OAM beams \cite{Marrucci2011,Marrucci2006}. Besides
their utility in a wide series of optical experiments, q-plates represent
indeed one of the most intriguing applications of the well-known Pancharatnam-Berry
phase phenomenon, presented in \cite{Berry1987}.

Several of the above mentioned techniques have been extended to the
radio frequency domain, where many other methods have also been proposed,
leading to an explosive growth in OAM-related publications (\cite{Barbuto2014,Cheng2014,Fonseca2015,Maccalli2013,Niemiec2014,Tennant2012,Yu2016,Zheng2016}, just to cite a few). As originally proposed in \cite{Thide2007},
antenna arrays can be employed to generate vortex radio waves upon
proper choice of the feeding currents/voltages. The use of uniform
circular arrays (UCAs\nomenclature{UCA}{uniform circular array}),
which consist of a circle of equally spaced radiators, probably represents
one of the best solutions, exploiting the natural vortex waves circular
symmetry for the benefit of a reduced number of antennas. Despite
these advantages, in this work it is shown that the circular symmetry
of the antenna array is not enough to attain good mode purity and
also the radiation and symmetry properties of each constituent element
must be taken into account. This, in turn, strongly impacts on the
efficiency of a vortex communication link based on UCAs, an issue that 
seems to have been rather neglected so far.

In Section \ref{sec:Generation}, the vortex waves generation by means
of UCAs is reviewed and the problem of symmetry breaking of the field
profiles is addressed with the help of several examples. The vortex
content of the channel modes relative to some communication links
between UCAs is explored in Section \ref{sec:channelMatrices}, where
the effect of symmetry breaking on the mode orthogonality is presented.
The most important results are briefly summarized in the conclusions.
As a support to the main document, four detailed appendices are devoted
to theoretical insights and additional calculations. 

\section{Generation of vortex waves via phased antenna arrays\label{sec:Generation}}

Antenna arrays are configurations of multiple connected antennas, distributed 
according to a specific geometry, in which the excitation of each radiator 
can be controlled in amplitude and phase \cite{Collin1969,Orfanidis2016}. 
As the electromagnetic waves radiated by the constituent
elements combine coherently, affecting the spatial distribution of the transmitted 
power, antenna arrays represent one of the most manageable and versatile
solutions to deal with the synthesis of complex waveforms at the radio
frequencies. Furthermore, unlike the majority of other systems involving
a single radiator, they are well suited for the simultaneous
transmission of several independent beams, each of which can be encoded
with a different information content and electronically controlled
in direction, thus ensuring a high flexibility for multiplexing \cite{Chandran2004,Haupt2010,Rabinovich2013}.
In the theoretical model that will be developed, the presence of an ideal
beamforming network which provides the array excitations is assumed:
noise, energy losses and couplings among the antennas will not be
considered. Actually, in view of a practical implementation, many
of the unbalances that may come into play in a realistic scenario
could be treated with the help of the electronics or digitally controlled
systems. 

\begin{figure}[t]
\noindent \begin{centering}
\includegraphics[width=1\textwidth]{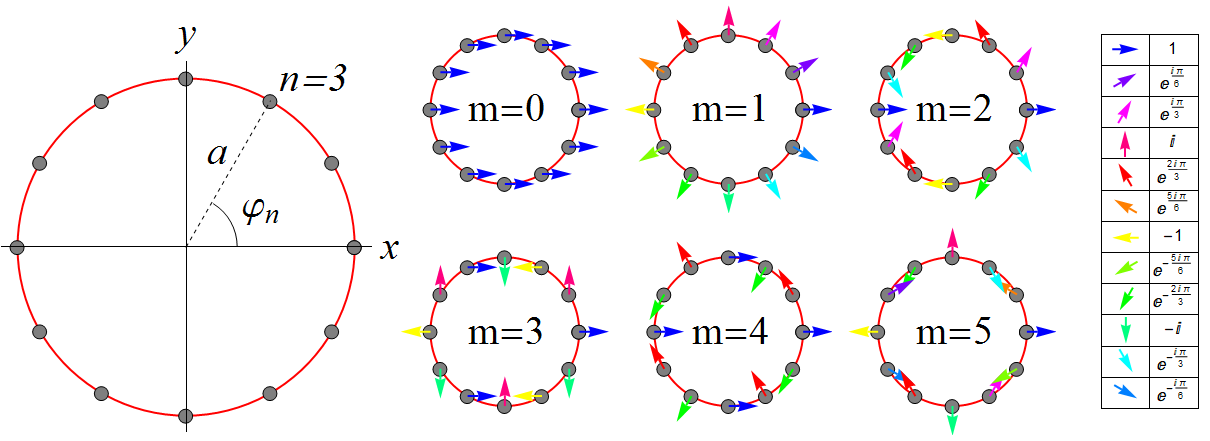}
\par\end{centering}
\caption{Graphical representation of a UCA composed of $N=12$ elements and
of the excitation coefficients $\varPhi_{m}^{n}$, defined in (\ref{eq:UCAcoeff}),
for some values of $m$.\label{fig:uca}}
\end{figure}

According to the literature, a UCA composed of $N$ arbitrary elements
can generate a vortex wave with topological charge $m$ in the integer
range $\left(-N/2,\:N/2\right)$ when the input excitation to the
$n$-th antenna terminals is expressed as:
\begin{equation}
\varPhi_{m}^{n}=\frac{1}{\sqrt{N}}\exp\left(im\varphi_{n}\right)=\frac{1}{\sqrt{N}}\exp\left[2\pi im\left(\frac{n-1}{N}\right)\right],\label{eq:UCAcoeff}
\end{equation}
being $\varphi_{n}$ the angular position of the $n$-th element of
the UCA, as shown in Figure \ref{fig:uca}. The set $\left\{ \varPhi_{m}^{n}\right\} $
can be interpreted as the discretized version of the function $\varPhi_{m}\left(\varphi\right)/\sqrt{2\pi}$,
where $\varPhi_{m}\left(\varphi\right)$ is the azimuthal phase term
characterizing vortex waves. Orthogonality and completeness relations
(to be compared with their continuous counterparts) are also derived:
\begin{equation}
\sum_{n=1}^{N}\left(\varPhi_{j}^{n}\right)^{*}\varPhi_{m}^{n}=\sum_{n=1}^{N}\frac{\exp\left[i\varphi_{n}\left(m-j\right)\right]}{N}=\delta_{jm};\quad\sum_{m}\varPhi_{m}^{n}\left(\varPhi_{m}^{p}\right)^{*}=\delta_{np},\label{eq:orthogonalityCompletenessDiscrete}
\end{equation}
where the azimuthal indices $m$ and $j$ are required to span a set
of $N$ integer values, that is $\left[-N/2,\:N/2\right)$ or $\left(-N/2,\:N/2\right]$
for $N$ even, in which case the extreme values $\pm N/2$ lead to
degenerate configurations with spurious vortex contribution\footnote{Sampling arguments can be used to show that for $m=\pm N/2$ and $N$
even the two corresponding phase weighting configurations $\left\{ \varPhi_{m}^{n}\right\} $
coincide and therefore cannot excite distinct fields. The pattern
relative to such limiting cases does not represent a pure vortex wave
but rather a superposition of the two.}, and $\left[-N/2,\:N/2\right]$ for $N$ odd. Both equations in (\ref{eq:orthogonalityCompletenessDiscrete})
directly follow from the summation formula of the geometric progression
to $N$ terms \cite{Abramowitz1972}.

Let us consider a UCA of $N$ ideal isotropic radiators lying in the
$xy$ plane; the frequency-domain analytic expression of the scalar
vortex wave generated by such distribution reads:
\begin{eqnarray}
f_{m}\left(\mathbf{r}\right) & = & \int d^{3}r'\:g_{+}\left(\mathbf{r}\text{\textminus}\mathbf{r}'\right)\sum_{n=1}^{N}\varPhi_{m}^{n}\delta^{3}\left(\mathbf{r}'-\mathbf{r}_{n}\right)\nonumber \\
 & = & -\sum_{n=1}^{N}\varPhi_{m}^{n}\frac{\exp\left(-ik\left|\mathbf{r}-\mathbf{r}_{n}\right|\right)}{4\pi\left|\mathbf{r}-\mathbf{r}_{n}\right|},\label{eq:campoUCAisotropi}
\end{eqnarray}
where $g_{+}$ is the outgoing-wave Green function defined in (\ref{eq:GreenFunction}), $m$ is the topological charge, $\left|m\right|<N/2$, $\mathbf{r}_{n}$
identifies the position of the $n$-th spherical wave source and all
dimensional constants have been neglected for simplicity. Equation
(\ref{eq:campoUCAisotropi}) is derived from (\ref{eq:radiatedField}) 
upon proper choice of the source term.
Although approximated closed-form expressions could be obtained from
(\ref{eq:campoUCAisotropi}) under specific assumptions (for example,
a very large number of array elements), these will not be taken into
consideration for brevity. Instead, we can resort to the useful far-field
approximation $r\gg a$, being $a$ the radius of the array, which
leads to: 
\begin{equation}
f_{m}\left(r,\theta,\phi\right)\approx-\frac{\exp\left(-ikr\right)}{4\pi\sqrt{N}r}\sum_{n=1}^{N}\exp\left[ika\sin\theta\cos\left(\phi-\varphi_{n}\right)+im\varphi_{n}\right],\label{eq:campoUCAisotropiFarfield}
\end{equation}
where $\left\{ r,\theta,\phi\right\} $ represents the spherical coordinate
system and the expression $\mathbf{r}_{n}=a\left\{\cos\varphi_{n},\sin\varphi_{n},0\right\}$
has been employed.

\begin{figure}[t]
\noindent \begin{centering}
\includegraphics[width=1\textwidth]{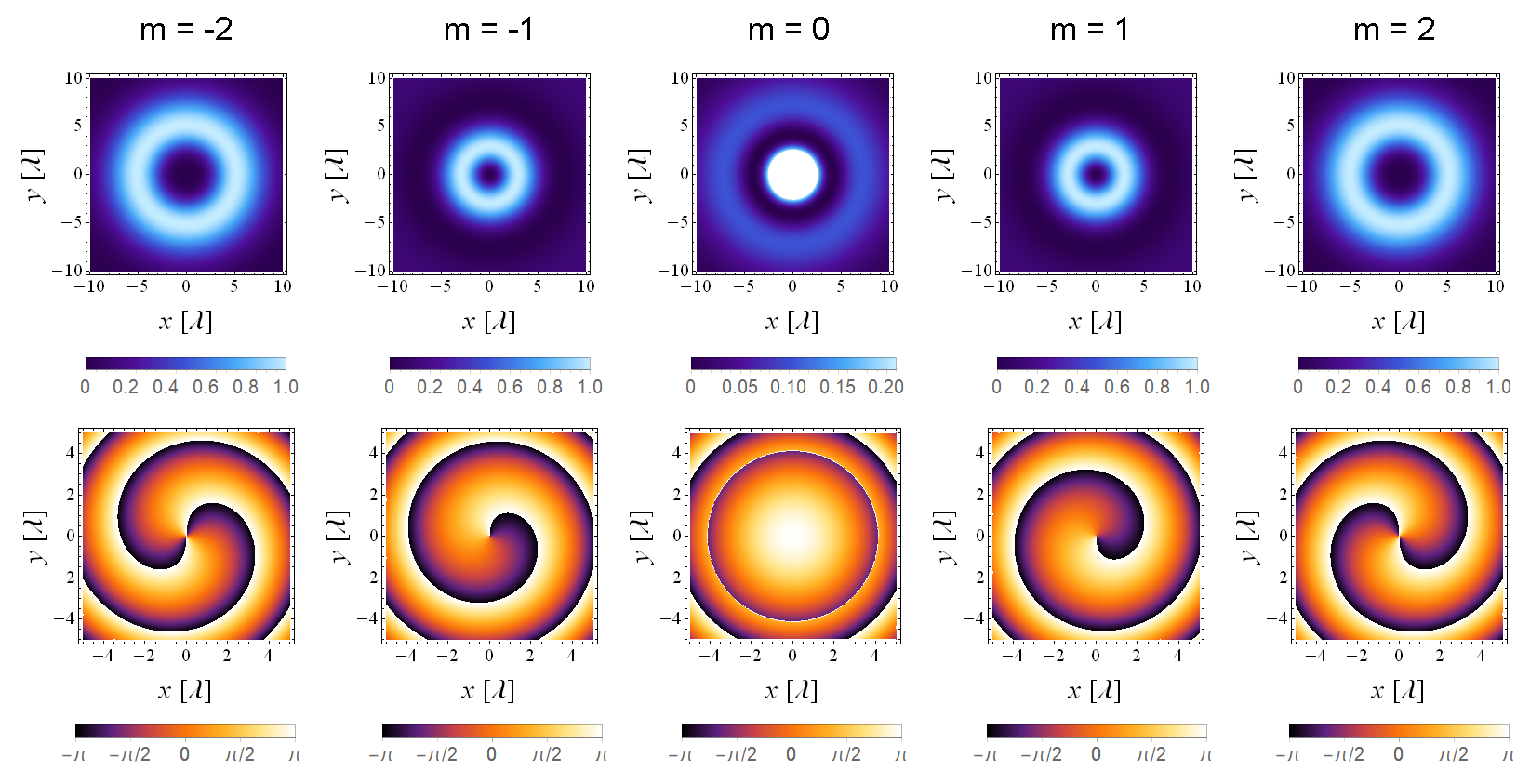}
\par\end{centering}
\caption{Relative intensity (upper row) and phase (lower row) profiles of the scalar
wave $f_{m}$ in (\ref{eq:campoUCAisotropiFarfield}) for $a=\lambda=1$
m, $N=12$ and $z=10\lambda$. To better highlight fainter details, the $m=0$ intensity maximum has been clipped.\label{fig:isoUCAfield}}
\end{figure}

Figure \ref{fig:isoUCAfield} displays some of the transverse intensity
and phase profiles of the function $f_{m}\left(\mathbf{r}\right)$
at varying $m$ in a fixed $z$-plane for a UCA like that reported
in Figure \ref{fig:uca}. The mode patterns are found to be in good
agreement with those expected for a scalar vortex solution but, of
course, neither (\ref{eq:campoUCAisotropi}) nor (\ref{eq:campoUCAisotropiFarfield})
suffices to describe a realistic vortex waveform.

Let us now move to the case of an arbitrary $N$-element antenna array
composed of more realistic radiators. The knowledge of the current
term $\mathbf{J}\left(\mathbf{r}\right)$ enables the estimation
of the electric and magnetic fields via (\ref{eq:electricGreenScalar})
and (\ref{eq:magneticGreenScalar}), from which the corresponding
far-field expressions are easily derived \cite{Orfanidis2016}. Neglecting
all mutual couplings among the antennas, the total electric field
radiated by the considered array in a region where $r$ is much greater
than the spatial extent of the antennas can be written as:
\begin{equation}
\mathbf{E}\left(r,\theta,\phi\right)=ik\eta I_{0}\sum_{n=1}^{N}\xi_{n}\frac{\exp\left(-ik\left|\mathbf{r}-\mathbf{r}_{n}\right|\right)}{4\pi\left|\mathbf{r}-\mathbf{r}_{n}\right|}\,\mathbf{h}_{n}\left(\theta,\phi\right),\label{eq:Earray}
\end{equation}
being $\eta$ the vacuum impedance, $I_{n}=I_{0}\,\xi_{n}$ the supply
current to the $n$-th radiator, where $\xi_{n}\in\mathbb{C}$ is
the relative synthesis coefficient and $I_{0}$ a constant current
term associated to the total input power, whereas $\mathbf{h}_{n}\left(\theta,\phi\right)$
represents the $n$-th antenna vector effective height (see \ref{sec:GreenFunctions}, equation (\ref{eq:effectiveHeight})). 

In deriving equation (\ref{eq:Earray}), no assumption was made on
the feeding, arrangement and type of antennas employed. By specializing
the array geometry to the case of a UCA in the $xy$ plane, it is
possible to analyze the properties of the electric field (\ref{eq:Earray})
associated with a set of vortex weighting coefficients $\xi_{n}=\varPhi_{m}^{n}$.
Under this hypothesis, and for $r\gg a$, the following far-field
expression holds:
\begin{equation}
\mathbf{E}_{m}\left(r,\theta,\phi\right)\approx\frac{ik\eta I_{0}\exp\left(-ikr\right)}{4\pi\sqrt{N}r}\sum_{n=1}^{N}\mathbf{h}_{n}\left(\theta,\phi\right)\mathrm{e}^{ika\sin\theta\cos\left(\phi-\varphi_{n}\right)+im\varphi_{n}}.\label{eq:EarrayFarfield}
\end{equation}

\begin{figure}[!t]
\noindent \begin{centering}
\includegraphics[width=1\textwidth]{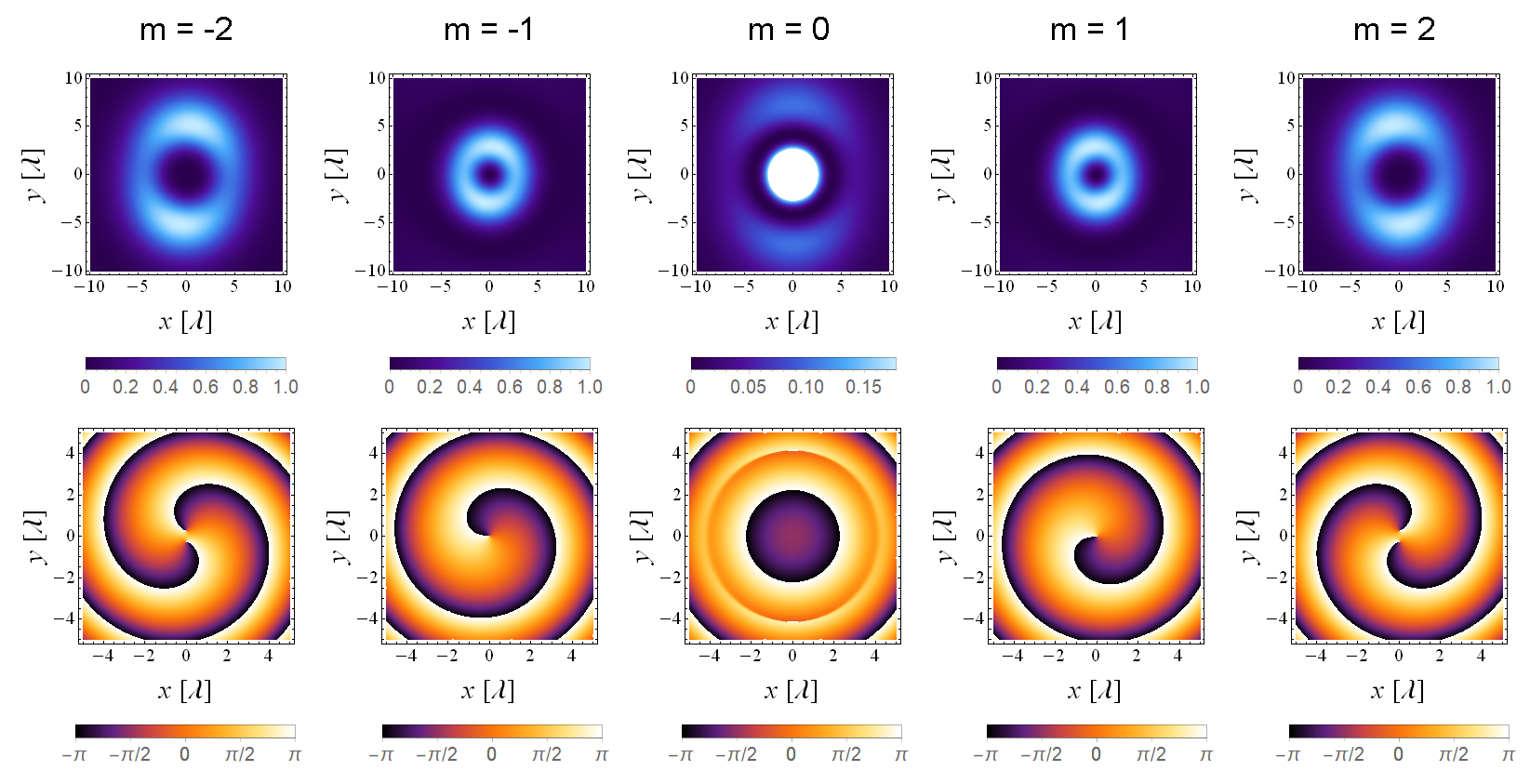}
\par\end{centering}
\caption{Relative intensity (upper row) and phase (lower row) profiles of the $x$-component
of the electric field (\ref{eq:EarrayFarfield}) radiated by a UCA
composed of $N=12$ $x$-polarized half-wave dipoles, for $a=\lambda=1$
m and $z=10\lambda$.\label{fig:linUCAfield}}
\end{figure}

Some of the transverse intensity and phase profiles of the $x$-component
of $\mathbf{E}_{m}\left(\mathbf{r}\right)$ computed
in the plane $z=10\lambda$ are reported in Figure \ref{fig:linUCAfield}
for the case of a UCA composed of $N=12$ $x$-directed half-wave
dipoles. It is fundamental to note how the use of linear sources 
breaks the degenerate symmetry of the isotropic case, causing
a deviation of the patterns from the expected ones. In particular,
the intensity distributions are no longer perfectly circular and the
phase profiles witness a corruption of the on-axis vortex with charge
$m$ in $m$ distinct off-axis vortices with unitary charge. 
\begin{figure}
\noindent \begin{centering}
\includegraphics[width=1\textwidth]{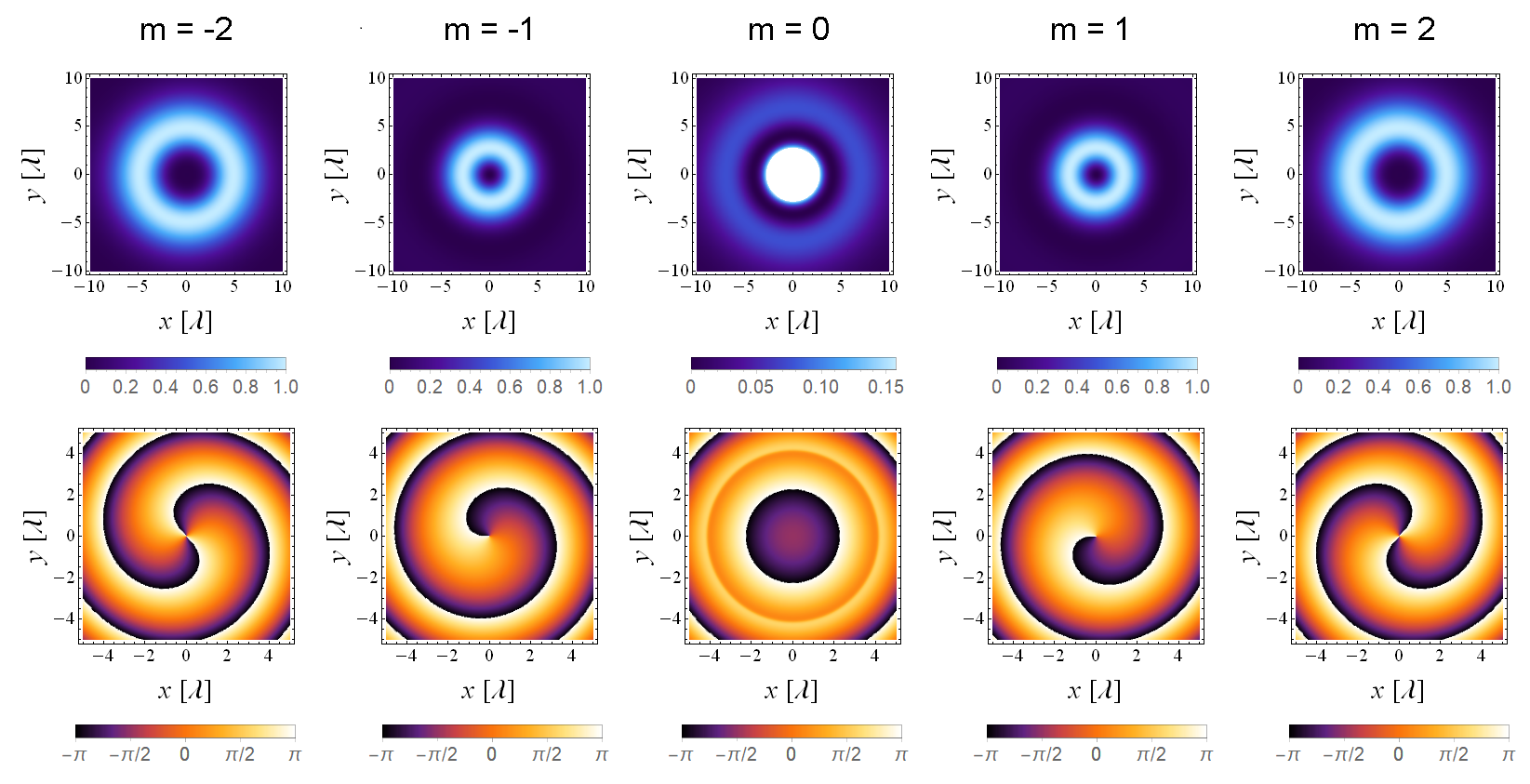}
\par\end{centering}
\caption{Relative intensity (upper row) and phase (lower row) profiles of the $\mathbf{u}_{x}-i\mathbf{u}_{y}$
component of the right circularly polarized electric field (\ref{eq:EarrayFarfield})
radiated by a UCA composed of $N=12$ crossed half-wave dipoles arranged
along the $x$ and $y$ directions, for $a=\lambda=1$ m and $z=10\lambda$.\label{fig:crosUCAfield}}
\end{figure}

\begin{figure}[H]
\noindent \begin{centering}
\includegraphics[width=1\textwidth]{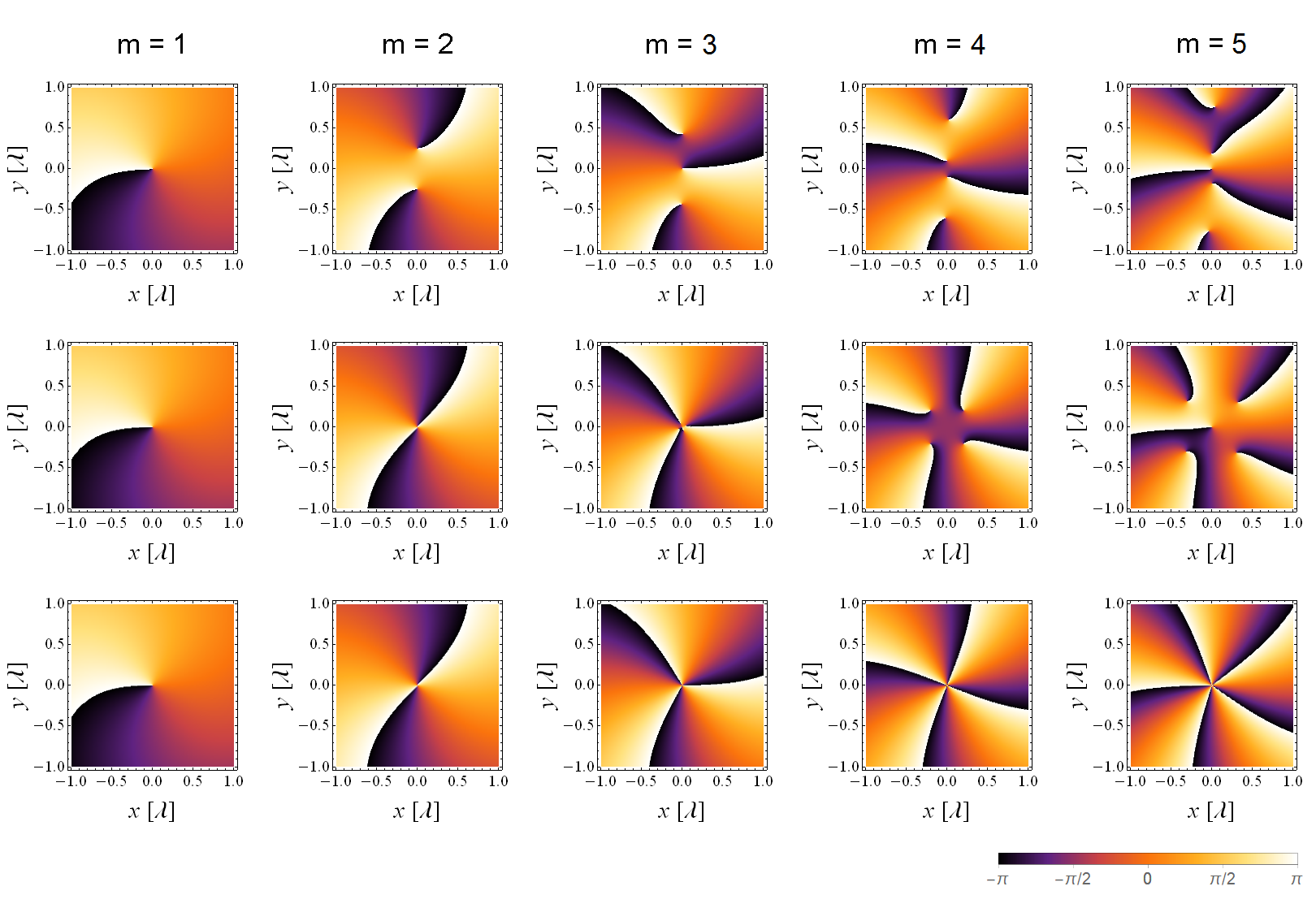}
\par\end{centering}
\caption{Phase profiles of the electric field (\ref{eq:EarrayFarfield}) radiated
by a UCA composed of $N=12$ half-wave dipoles ($x$-component, first
row), right circularly polarized crossed half-wave dipoles ($\mathbf{u}_{x}-i\mathbf{u}_{y}$
component, second row) and right circularly polarized crossed Hertzian
dipoles of length $\lambda/20$ ($\mathbf{u}_{x}-i\mathbf{u}_{y}$ component, third
row), for $a=\lambda=1$ m and $z=10\lambda$.\label{fig:phaseComparison}}
\end{figure}

As shown in Figure \ref{fig:crosUCAfield} and in Figure \ref{fig:phaseComparison},
the standard vortex profiles can be recovered by using low-directive circular symmetric
array elements such as crossed dipoles with a $\pi/2$ phase delay,
in which case also a spin angular momentum (SAM) contribution will
be present. It is worth emphasizing that circular polarization alone is not enough to 
completely solve the symmetry breaking of vortex waves, as confirmed by the fact 
that higher order profiles obtained with crossed Hertzian dipoles are better 
than those produced by crossed half-wave dipoles, the former type of antennas  
being closer to the ideal spherical wave source.

More generally, owing to the vector nature of the above equations,
it is possible to reproduce approximate versions of the vector vortex
solutions \cite{VolkeSepulveda2002}. For instance, the azimuthally and radially polarized vector
beams are found by considering UCAs of dipoles arranged in the $\mathbf{u}_{\varphi}$-direction
and in the $\mathbf{u}_{\rho}$-direction, respectively\footnote{It is interesting to note that such configurations belong to the class
of the so-called \textit{ring quasi-arrays}, originally proposed long
ago and analyzed in connection with their supposed super-gain properties
\cite{Knudsen1956}.}. As an alternative, any physical rotation of the dipoles in the circumference
plane can be replaced by a rotation of the current vectors implemented
through a UCA of properly fed crossed dipoles displaced along two
fixed orthogonal directions.

\begin{figure}[!t]
\noindent \begin{centering}
\includegraphics[width=1\textwidth]{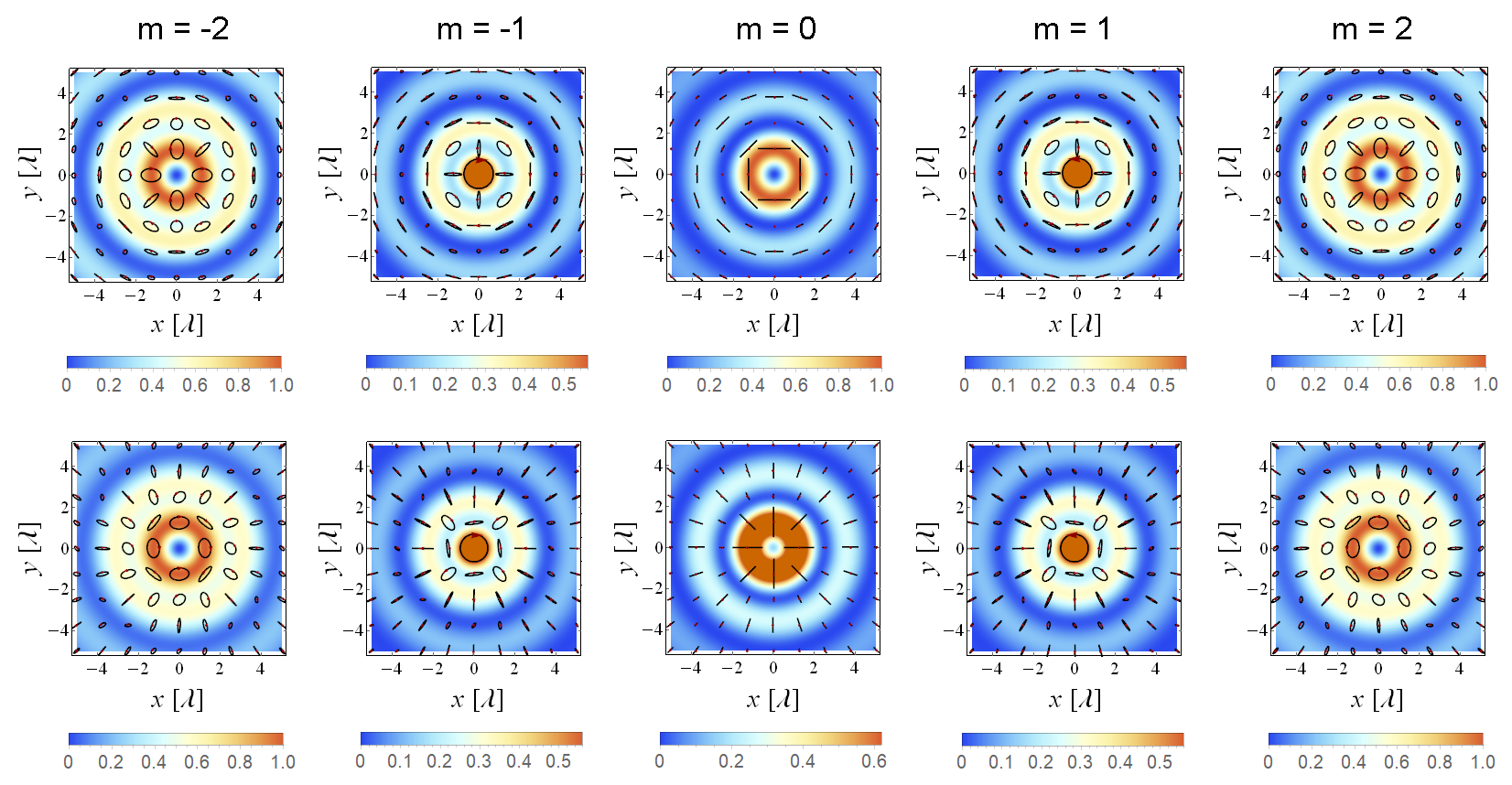}
\par\end{centering}
\caption{Norm and polarization plots of the electric field (\ref{eq:EarrayFarfield})
radiated by a UCA composed of $N=25$ half-wave dipoles arranged along
the $\mathbf{u}_{\varphi}$-direction (upper row) and the $\mathbf{u}_{\rho}$-direction
(lower row), for $a=1$ m, $\lambda=0.4$ m and $z=10\lambda$. Some of the intensity maxima have been clipped for better clarity.\label{fig:aziRadUCAprofiles}}
\end{figure}

The electric field norm and polarization plots of some vector waves
generated by UCAs of azimuthally and radially polarized half-wave
dipoles are displayed in Figure \ref{fig:aziRadUCAprofiles}, showing
some of the interesting features that characterize the theoretical
vector vortex solutions. Not surprisingly, the possibility of synthesizing
vector vortex beams by means of circular arrays of sources oriented
along $\mathbf{u}_{\varphi}$ and $\mathbf{u}_{\rho}$ has
also been explored in optics as a theoretical tool for modeling optical
vortex beams emitters \cite{Zhu2013,Zhu2014}.

As a further point, one may wonder if it is possible to generalize
the set $\left\{ \varPhi_{m}^{n}\right\} $ in (\ref{eq:UCAcoeff})
beyond the UCA case without the need for a complete array synthesis
procedure (that usually requires to solve a linear inverse problem).
For instance, the new set of vortex excitation coefficients could
be heuristically defined via sampling of some orthogonal basis functions
which contain the sought term $\exp\left(im\varphi\right)$ and are
supported on the reference volume for the transmitting array. Limiting
to the case of a uniform disk antenna array of radius $a$, a complete
set of functions that are orthogonal in the interior of the disk is
provided by the well-known generalized Zernike polynomials \cite{Born1999}:
\begin{equation}
Z_{q}^{m}\left(\varrho,\varphi\right)=R_{q}^{m}\left(\varrho\right)\exp\left(im\varphi\right),
\end{equation}
where $\varrho=\rho/a$ represents the scaled radial coordinate, $m$
is the azimuthal index, $q\leq\left|m\right|$ corresponds to a non-negative
integer such that $q-\left|m\right|$ is even and the radial function
is expressed by:
\begin{equation}
R_{q}^{m}\left(\varrho\right)=\sum_{s=0}^{\frac{q-\left|m\right|}{2}}\frac{\left(-1\right)^{s}\left(q-s\right)!}{s!\left(\frac{q+\left|m\right|}{2}-s\right)!\left(\frac{q-\left|m\right|}{2}-s\right)!}\varrho^{q-2s}.
\end{equation}
For each of the $N$ antennas regularly distributed over the transmitting
disk, $Z_{q}^{m}\left(\varrho_{n},\varphi_{n}\right)$ identifies
the local value of the Zernike function with indices $q$ and $m$
and can be used to build a corresponding column vector of vortex excitations
for the array. It should be remembered that the discretization introduces
bounds on the possible values of the azimuthal and radial indices
and limits the number of effectively orthogonal column vectors. 
\begin{figure}[!t]
\noindent \begin{centering}
\includegraphics[width=1\textwidth]{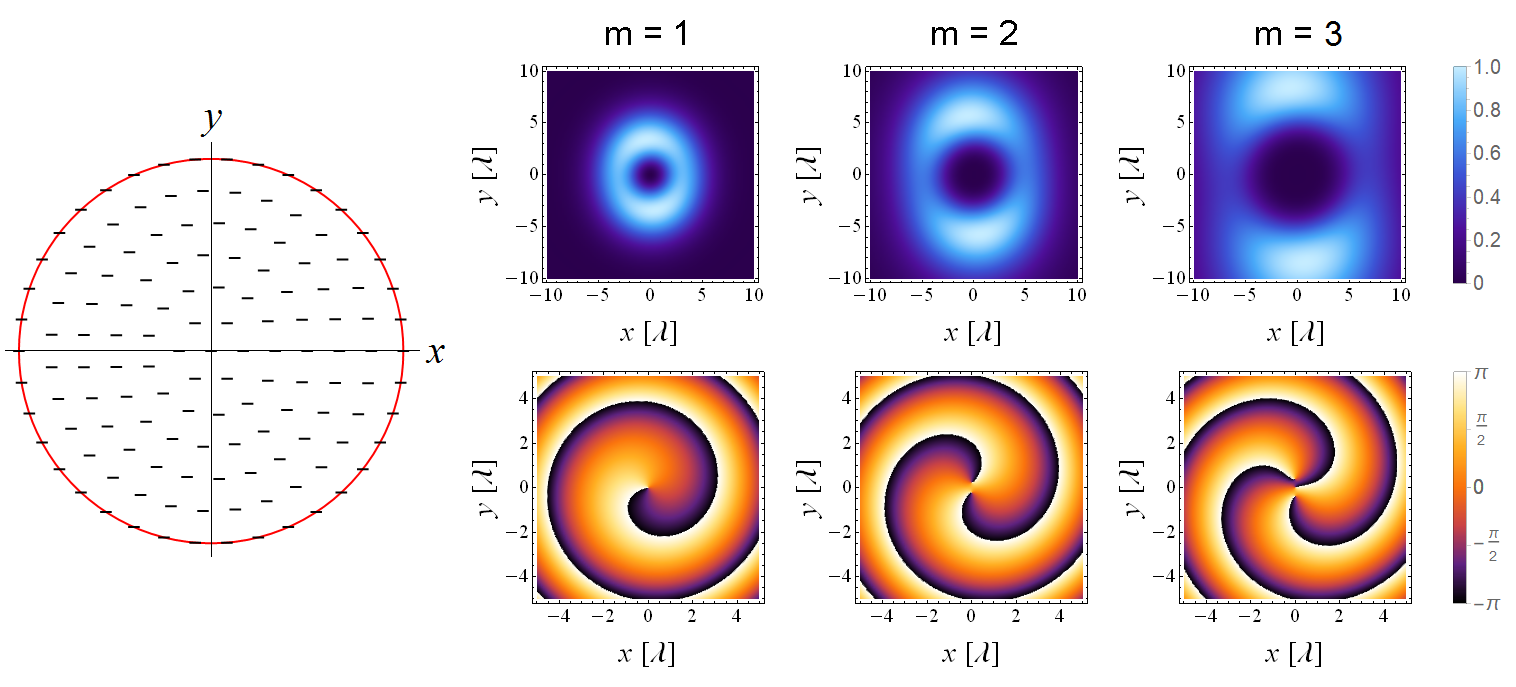}
\par\end{centering}
\caption{Synthesis of some vortex waves via sampling of the Zernike polynomials
on a uniform disk array of $x$-polarized Hertzian dipoles of length $\lambda/20$: sketch
of the geometry (left), relative intensity and phase profiles of the $x$-component
of the synthesized electric field evaluated at $z=10\lambda$ (right),
for $N=133$ and $a=\lambda=1$ m.\label{fig:zernikeSynthesisLin}}
\end{figure}
\begin{figure}[!t]
\noindent \begin{centering}
\includegraphics[width=1\textwidth]{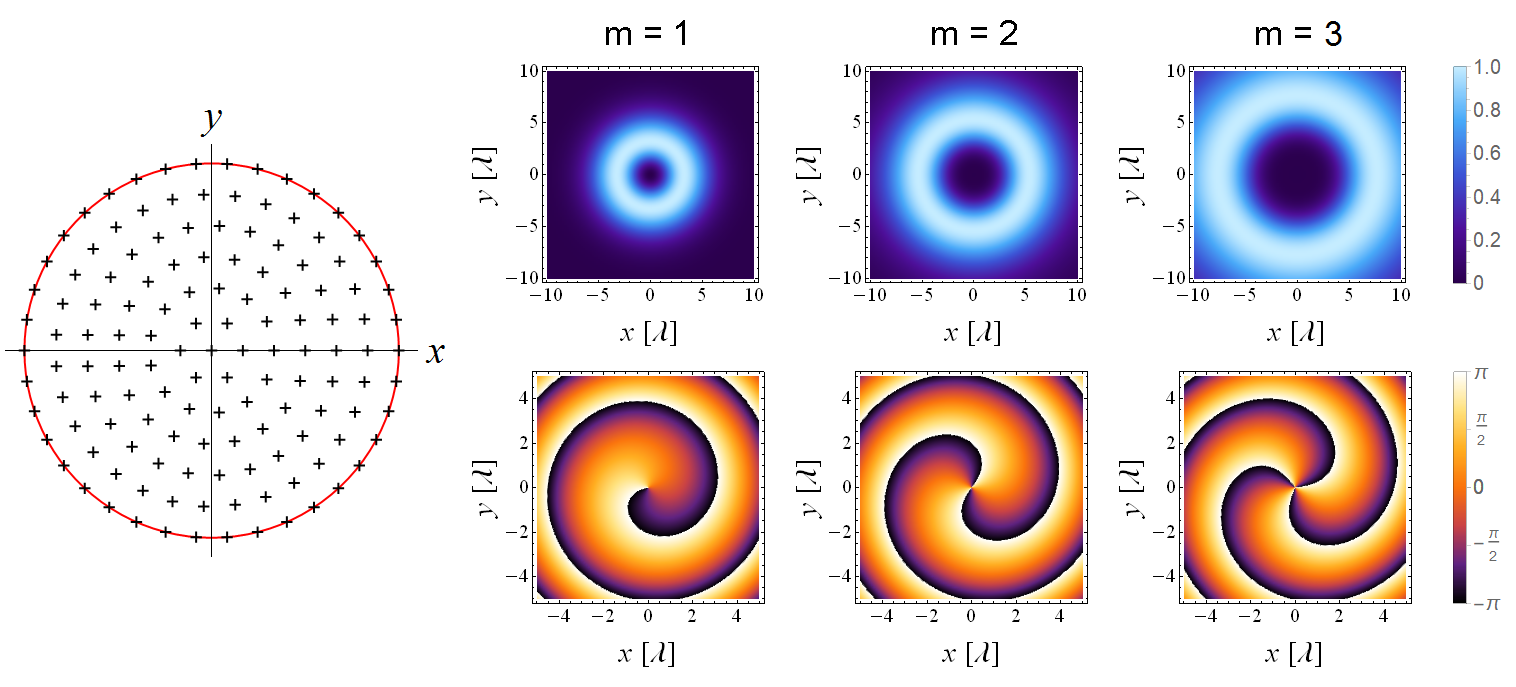}
\par\end{centering}
\caption{Synthesis of some vortex waves via sampling of the Zernike polynomials
on a uniform disk array of right circularly polarized crossed Hertzian
dipoles of length $\lambda/20$: sketch of the geometry (left), relative intensity and phase profiles
of the $\mathbf{u}_{x}-i\mathbf{u}_{y}$ component of the
synthesized electric field evaluated at $z=10\lambda$ (right), for
$N=133$ and $a=\lambda=1$ m.\label{fig:zernikeSynthesisCros}}
\end{figure}
For the sake of simplicity, let us neglect the radial degrees of freedom
by fixing $q=\left|m\right|$ and consider the set made of the $N$-length
column vectors $\left\{ Z_{\left|m\right|}^{m}\left(\varrho_{n},\varphi_{n}\right)\right\} $.

Figure \ref{fig:zernikeSynthesisLin} and Figure \ref{fig:zernikeSynthesisCros}
display the electric field profiles relative to some of the effectively
orthogonal sampled coefficients for a disk array composed of uniformly
spaced $x$-directed Hertzian dipoles and circularly polarized crossed
Hertzian dipoles, respectively. The results are easily interpreted
in accordance with what seen above and could be generalized to other
geometries.

\section{Basis sets and circular symmetry\label{sec:channelMatrices}}

As suggested by David A. B. Miller in \cite{Miller2000}, the orthogonal
spatial information channels, also referred to as the communications
modes or simply the channel modes, can be interpreted as the pairs
of functions that diagonalize the mathematical operator formalizing
the communications between the transmitting and the receiving volumes.
In \ref{sec:Communication}, the abstract operator is specialized
to the case of a finite-dimensional channel matrix which describes
the communications between two arbitrary antenna arrays. The same
concepts are applied to some specific examples in the present section.
Within the proposed framework, the singular vectors of the channel
matrix are analyzed in relation to the standard vortex modes basis,
shedding light onto some of the interesting features that arise from
the intimate connection between system geometry and symmetry properties
of the channel modes. 

In the following, transmitting and receiving arrays are considered
identical in spatial disposition, number of elements, labeling order and type
of constituent antennas, for simplicity. Under this assumption, the
Cartesian coordinates $\mathbf{r}_{p}^{R}=\left\{ x_{p}^{R},y_{p}^{R},z_{p}^{R}\right\} $
defining the locations of the receiving elements are linked to those
relative to the transmitting ones $\mathbf{r}_{n}^{T}=\left\{ x_{n}^{T},y_{n}^{T},z_{n}^{T}\right\} $
via the matrix equation:
\begin{equation}
\mathbf{r}_{p}^{R}=\mathsf{R}\,\mathbf{r}_{p}^{T}+\mathsf{\mathbf{t}},\label{eq:rototranslation}
\end{equation}
where $\mathsf{R}$ represents a three-dimensional rotation matrix
and $\mathsf{\mathbf{t}}$ a translation vector (for instance,
$\mathsf{R}=\mathbb{I}$ and $\mathsf{\mathbf{t}}=\left\{ 0,0,d\right\} $
in the case of perfectly aligned antenna arrays separated by a distance
$d$). From simple geometrical arguments, the set of unit vectors
$\mathbf{v}=\left\{ \mathbf{u}_{x},\mathbf{u}_{y},\mathbf{u}_{z}\right\} $
employed to describe the antenna orientation transforms accordingly:
\begin{equation}
\mathbf{v}_{p}^{R}=\mathsf{R}^{-1}\mathbf{v}_{p}^{T}.\label{eq:rototranslationVectors}
\end{equation}

As a straightforward example, let $\mathcal{V}_{T}$ and $\mathcal{V}_{R}$
be two identical facing circumferences of radius $a$ at a distance
$d$ apart and suppose two UCAs composed of $N$ $x$-polarized Hertzian
dipoles of length $l$ to be arranged along such circumferences. Since
the effective height of a $z$-directed Hertzian antenna placed in
the origin reads $l\sin\theta\,\mathbf{u}_{\theta}$, the channel
matrix (\ref{eq:channelMatrix}) for this aligned communication system is simply given by:
\begin{equation}
H_{pn}=ik\eta I_{0}\frac{\exp\left(-ik\left|\mathbf{r}_{p}^{R}-\mathbf{r}_{n}^{T}\right|\right)}{4\pi\left|\mathbf{r}_{p}^{R}-\mathbf{r}_{n}^{T}\right|}\left(l\sin\theta_{np}^{x}\right)^{2},\label{eq:channelMatrixHertzian}
\end{equation}
being $\theta_{np}^{x}$ the angle between the vector $\left(\mathbf{r}_{p}^{R}-\mathbf{r}_{n}^{T}\right)$
and the $x$-axis. On using the explicit expressions for $\mathbf{r}_{p}^{R}$
and $\mathbf{r}_{n}^{T}$, equation (\ref{eq:channelMatrixHertzian})
reduces to:
\begin{equation}
H_{pn}=ik\eta I_{0}\frac{\exp\left(-ik\sqrt{d^{2}+2a^{2}\left\{ 1-\cos\left[\frac{2\pi\left(p-n\right)}{N}\right]\right\} }\right)}{4\pi\sqrt{d^{2}+2a^{2}\left\{ 1-\cos\left[\frac{2\pi\left(p-n\right)}{N}\right]\right\} }}\left(l\sin\theta_{np}^{x}\right)^{2}.\label{eq:channelMatrixHerzianOffset}
\end{equation}

\begin{figure}[t]
\noindent \begin{centering}
\includegraphics[width=1\textwidth]{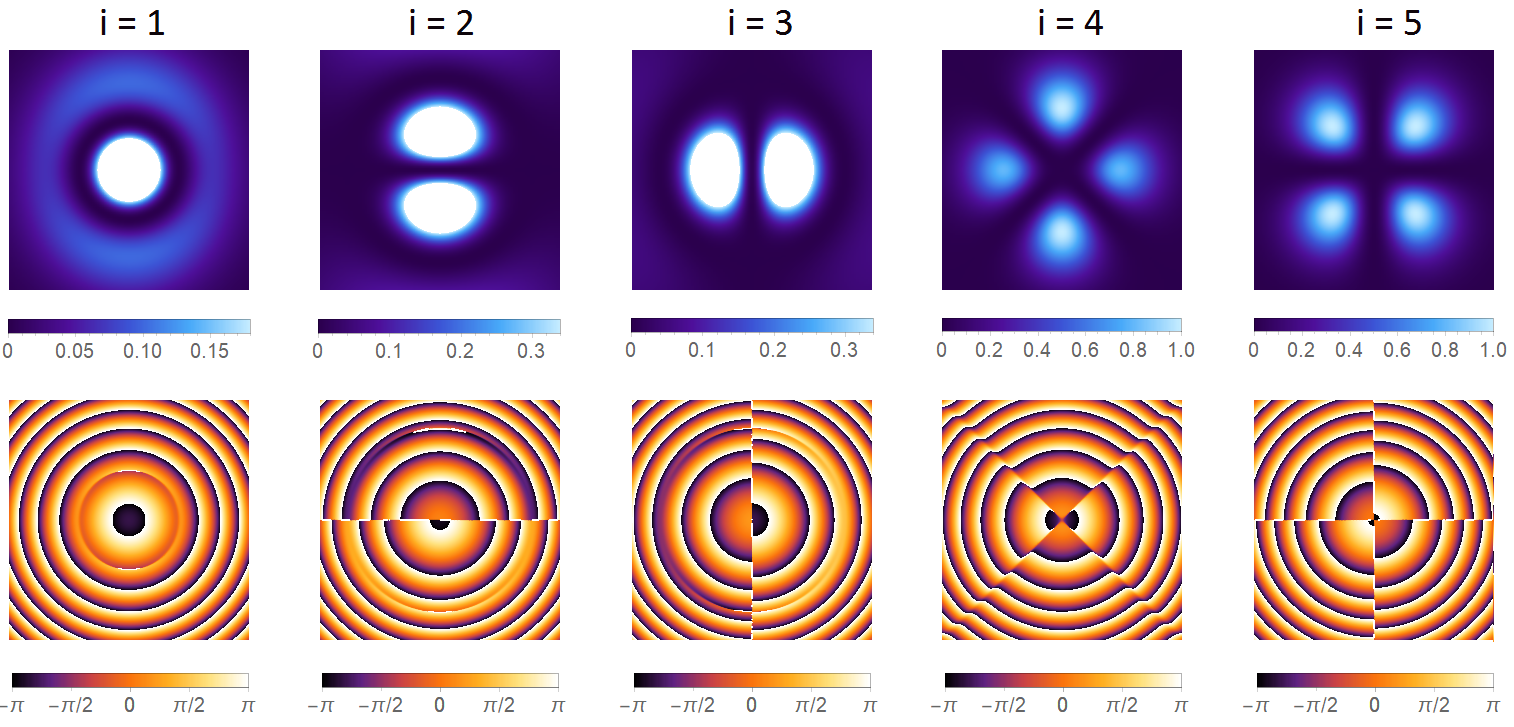}
\par\end{centering}
\caption{Norm (upper row) and phase ($x$-component, lower row) profiles of
the electric field radiated by a UCA composed of $N=25$ $x$-polarized
Hertzian dipoles with excitation coefficients given by the right singular
vectors $\left\{ \xi_{n}^{(i)}\right\} $ of the channel matrix (\ref{eq:channelMatrixHerzianOffset}),
for $a=\lambda=1$ m, $l=\lambda/20$, $d=10\lambda$. A clipping of the first three intensity maxima has been introduced to avoid faint plots. \label{fig:singularVectorsHertzian}}
\end{figure}

In Figure \ref{fig:singularVectorsHertzian} are reported some density
plots of the electric field radiated in the transverse plane $z=d=10\lambda$
by the transmitting UCA when the first five right singular vectors
of the channel matrix (\ref{eq:channelMatrixHerzianOffset}) are employed
as excitation coefficients.
 Analogous profiles can be found for the left singular vectors. Despite
no one of the wavefields excited by the considered channel modes presents
vortex character, some important subtleties are involved and will
now be discussed. 

In order to minimize the notation, let $\left|m\right\rangle =\left\{ \varPhi_{m}^{n}\right\} $
be a column vector in $\mathbb{C}^{N}$ and suppose the azimuthal
index value $m$ to be specified via the following rule:
\begin{equation}
m_{j}=\begin{cases}
-\frac{N}{2}+j-1 & N\:\mathrm{even};\\
-\frac{N-1}{2}+j-1 & N\:\mathrm{odd},
\end{cases}\label{eq:mConvention}
\end{equation}
with $j$ running from $1$ to $N$. By this convention, the collection
$\left\{ \left|m_{j}\right\rangle \right\} $ provides a basis set
for the space of all the possible UCA excitations $\left|\xi\right\rangle =\left\{ \xi_{n}\right\} $
and the orthogonality and completeness relations (\ref{eq:orthogonalityCompletenessDiscrete})
can be rewritten in Dirac notation:
\begin{equation}
\left\langle m_{i}\right.\left|m_{j}\right\rangle =\delta_{ij};\quad\sum_{j=1}^{N}\left|m_{j}\right\rangle \left\langle m_{j}\right|=\mathbb{I},\label{eq:DiracOrthogonalityCompleteness}
\end{equation}
where $\mathbb{I}$ is the $N\times N$ identity matrix. Now, labeling
the $i$-th right singular vector of the channel matrix with $\left|e_{i}\right\rangle =\left\{ \xi_{n}^{(i)}\right\} $,
the scalar product: 
\begin{equation}
c_{ji}=\left\langle m_{j}\right.\left|e_{i}\right\rangle \label{eq:spectralCoefficient}
\end{equation}
can be understood as the $j$-th expansion coefficient of $\left|e_{i}\right\rangle $
over the vortex modes basis $\left\{ \left|m_{j}\right\rangle \right\} $,
namely: 
\begin{equation}
\left|e_{i}\right\rangle =\sum_{j=1}^{N}\left|m_{j}\right\rangle \left\langle m_{j}\right.\left|e_{i}\right\rangle =\sum_{j=1}^{N}c_{ji}\left|m_{j}\right\rangle ,\label{eq:spectralExpansion}
\end{equation}
as follows from the completeness condition. 

\begin{figure}[!t]
\noindent \begin{centering}
\includegraphics[width=1\textwidth]{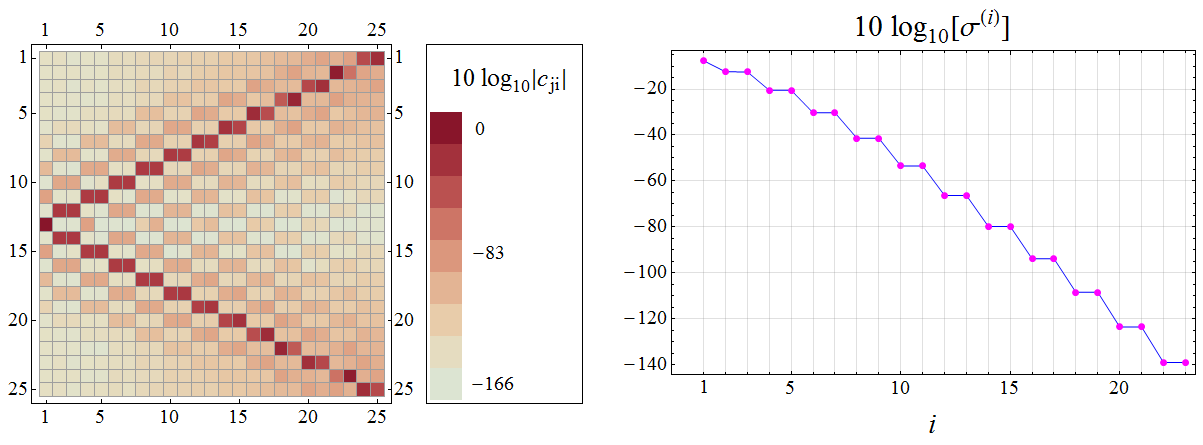}
\par\end{centering}
\caption{Spectral analysis of the channel operator (\ref{eq:channelMatrixHerzianOffset}):
matrix plot of the scalar product $c_{ji}$ defined in (\ref{eq:spectralCoefficient})
(left) and singular values of $H_{pn}$ as a function of the spectral
index (right), for $N=25$, $a=\lambda=1$ m, $l=\lambda/20$, $d=10\lambda$.\label{fig:matrixPlotHertzian}}
\end{figure}

\begin{figure}[!t]
\noindent \begin{centering}
\includegraphics[width=1\textwidth]{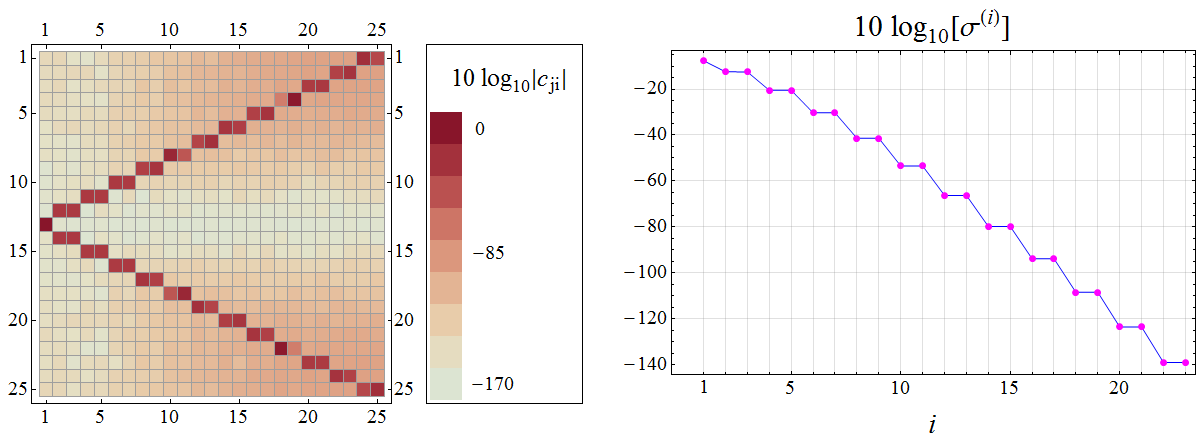}
\par\end{centering}
\caption{Spectral analysis of the channel operator (\ref{eq:channelMatrixIsotropic}):
matrix plot of the scalar product $c_{ji}$ (left) and singular values
of $H_{pn}^{(\circ)}$ as a function of the spectral index (right),
for $N=25$, $a=\lambda=1$ m, $l=\lambda/20$, $d=10\lambda$.\label{fig:matrixPlotIsotropic}}
\end{figure}

Figure \ref{fig:matrixPlotHertzian} (left) displays the projection
(\ref{eq:spectralCoefficient}) for the right singular vectors of
the channel matrix (\ref{eq:channelMatrixHerzianOffset}), highlighting
the presence of spurious vortex contributions in the matrix spectrum.
 This circumstance can be traced back to the use of non-isotropic
antennas, as clarified by considering a simplified version of (\ref{eq:channelMatrixHerzianOffset}):
\begin{equation}
H_{pn}^{(\circ)}=ik\eta I_{0}l^{2}\frac{\exp\left(-ik\sqrt{d^{2}+2a^{2}\left\{ 1-\cos\left[\frac{2\pi\left(p-n\right)}{N}\right]\right\} }\right)}{4\pi\sqrt{d^{2}+2a^{2}\left\{ 1-\cos\left[\frac{2\pi\left(p-n\right)}{N}\right]\right\} }},\label{eq:channelMatrixIsotropic}
\end{equation}
where the directivity of the arrays elements has been neglected and
the label $^{(\circ)}$ means ``isotropic case''. The projection
of the right singular vectors relative to (\ref{eq:channelMatrixIsotropic})
on the vortex modes basis is reported in Figure \ref{fig:matrixPlotIsotropic}
(left), from which it is apparent that each singular vector represents
an exact superposition of two vortex modes with opposite topological
charges, i.e. $\pm m$.

As it is clear from both Figure \ref{fig:matrixPlotHertzian} and
Figure \ref{fig:matrixPlotIsotropic}, the singular values of the
channel matrix for the transmission between facing UCAs are two-fold
degenerate due to the fact that, when resolved in terms of the vortex
modes, there is no difference between two configurations with opposite
$m$ (more on this in \ref{sec:degeneracy}). In the isotropic
case (\ref{eq:channelMatrixIsotropic}), pure vortex modes can be
directly obtained by implementing a rotation of the singular vectors
in their corresponding eigenspaces so as to select one of the two
directions along which there is no coexistence of opposite azimuthal
indices; in this sense, vortex modes constitute a legitimate basis
of singular vectors. This is not surprising, since it has been shown
that $H_{pn}^{(\circ)}$ is diagonalized by the discrete Fourier transform
(DFT\nomenclature{DFT}{discrete Fourier transform}) matrix $T_{pi}=\varPhi_{m_{i}}^{p}$,
as it only depends on the difference $\left(p-n\right)$,
and is therefore circulant \cite{Edfors2012}. 

While ensuring a pure vortex spectrum of the channel matrix, the removal
of the antennas directivity is nonphysical and does not prevent that
the radiated vortex waves still reveal some of the inaccuracies shown
in Figure \ref{fig:linUCAfield}, unless also the polarization constraint
is removed (in which case the electric field must be replaced by a
less realistic scalar function). 

The above considerations naturally lead to question whether it is
possible to get results close to that obtained for the isotropic case
without compromising antennas directivity and vector character of
the fields. When complex array layouts are involved, synthesis techniques
must be employed to determine a proper set of beamforming coefficients
able to yield approximations of the sought vortex fields; in this
context, the synthesized set might not represent a basis of singular
vectors (especially whenever the circular symmetry is not fully guaranteed)
or, if it were, there could be no reason to prefer the vortex set
over any other. On the contrary, the possibility of relying on the
well-known \textit{a priori} set of coefficients $\left\{ \varPhi_{m}^{n}\right\} $
as the very basis which diagonalizes the channel matrix in a scenario
involving UCAs would be of great benefit for some applications, where
it may bring technological simplifications. 

To generate pure vortex beams when only simple UCAs are concerned,
we replace the $x$-directed elements in the system above with circularly
polarized crossed dipoles that provide a major degree of rotational
symmetry. Assuming the dipoles of each pair to be oriented along the
$x$-axis and the $y$-axis, respectively, the channel matrix can
be written as the scalar product:
\begin{equation}
H_{pn}^{(\times)}=\left(\frac{\mathbf{E}_{n}^{x}+s_{T}\mathbf{E}_{n}^{y}}{\sqrt{2}}\right)\cdot\left(\frac{\mathbf{h}_{p}^{x}+s_{R}\mathbf{h}_{p}^{y}}{\sqrt{2}}\right),\label{eq:crossedDipolesMatrix}
\end{equation}
with $\mathbf{E}_{n}^{x}$ and $\mathbf{E}_{n}^{y}$ representing
the electric fields radiated by the $x$-directed and by the $y$-directed
elements of the $n$-th pair, respectively, and $\mathbf{h}_{p}^{x}$,
$\mathbf{h}_{p}^{y}$ the effective heights of the $p$-th receiving
pair. In (\ref{eq:crossedDipolesMatrix}), where the label $^{(\times)}$
stands for ``crossed dipoles case'', the complex term $s_{T}$ corresponds
to the ratio between the supply currents of the $y$-polarized and
$x$-polarized elements, which is supposed to be the same for all
the $N$ pairs; similarly, a further term $s_{R}$ has been introduced
in the receiving pairs which accounts for a possible beamforming phase
shift between the signals collected along the two orthogonal directions.
A fifty-fifty power splitting between the two elements of each pair
is enforced at both the transmitting and the receiving sides through
the factors $1/\sqrt{2}$. When $s_{T}=\pm i$, a circularly polarized
electric field is generated along the propagation axis and the choice
$s_{R}=s_{T}^{*}$ correctly implements the required chirality inversion
upon reception. 

For the considered case of facing UCAs, equation (\ref{eq:crossedDipolesMatrix})
reduces to:
\begin{eqnarray}
H_{pn}^{(\times)} & = & \left[\left(\sin\theta_{np}^{x}\right)^{2}+s_{T}s_{R}\left(\sin\theta_{np}^{y}\right)^{2}+\sin\theta_{np}^{x}\sin\theta_{np}^{y}\left(s_{T}+s_{R}\right)\boldsymbol{\theta}_{np}^{x}\cdot\boldsymbol{\theta}_{np}^{y}\right]\nonumber \\
 & \times & ik\eta\frac{I_{0}}{2}l^{2}\frac{\exp\left(-ik\sqrt{d^{2}+2a^{2}\left\{ 1-\cos\left[\frac{2\pi\left(p-n\right)}{N}\right]\right\} }\right)}{4\pi\sqrt{d^{2}+2a^{2}\left\{ 1-\cos\left[\frac{2\pi\left(p-n\right)}{N}\right]\right\} }},\label{eq:channelMatrixCrossed}
\end{eqnarray}
where the following conventions have been introduced:
\begin{eqnarray}
\mathbf{r}_{np} & = & \left\{ x_{np},y_{np},z_{np}\right\} =\mathbf{r}_{p}^{R}-\mathbf{r}_{n}^{T};\\
\theta_{np}^{x} & = & \arccos\left(x_{np}/r_{np}\right);\quad\phi_{np}^{x}=\arctan\left(z_{np}/y_{np}\right);\label{eq:sphericalX}\\
\boldsymbol{\theta}_{np}^{x} & = & \left\{ -\sin\theta_{np}^{x},\cos\theta_{np}^{x}\cos\phi_{np}^{x},\cos\theta_{np}^{x}\sin\phi_{np}^{x}\right\} ;\\
\theta_{np}^{y} & = & \arccos\left(y_{np}/r_{np}\right);\quad\phi_{np}^{y}=\arctan\left(x_{np}/z_{np}\right);\\
\boldsymbol{\theta}_{np}^{y} & = & \left\{ \cos\theta_{np}^{y}\sin\phi_{np}^{y},-\sin\theta_{np}^{y},\cos\theta_{np}^{y}\cos\phi_{np}^{y}\right\} .\label{eq:polarVectorlY}
\end{eqnarray}
Expressions (\ref{eq:sphericalX}) to (\ref{eq:polarVectorlY}) identify
the spherical coordinates and polar unit vectors relative to $\mathbf{r}_{np}$
in the reference frames fixed to the $x$-directed and $y$-directed
element of the $n$-th transmitting pair and oriented accordingly;
under relabeling $n\leftrightarrow p$, the same expressions are used
to represent the vector $\mathbf{r}_{pn}=-\mathbf{r}_{np}$
in the reference frames relative to the $p$-th receiving pair. It
is important to emphasize that equation (\ref{eq:channelMatrixCrossed})
is correct only under the assumption that the two UCAs are perfectly
aligned, in which case $\theta_{pn}^{x;y}=\pi-\theta_{np}^{x;y}$,
$\phi_{pn}^{x;y}=\pi+\phi_{np}^{x;y}$, $\boldsymbol{\theta}_{pn}^{x;y}=\boldsymbol{\theta}_{np}^{x;y}$,
$\boldsymbol{\theta}_{np}^{x;y}\cdot\boldsymbol{\theta}_{np}^{x;y}=1$. 

\begin{figure}[!t]
\noindent \begin{centering}
\includegraphics[width=1\textwidth]{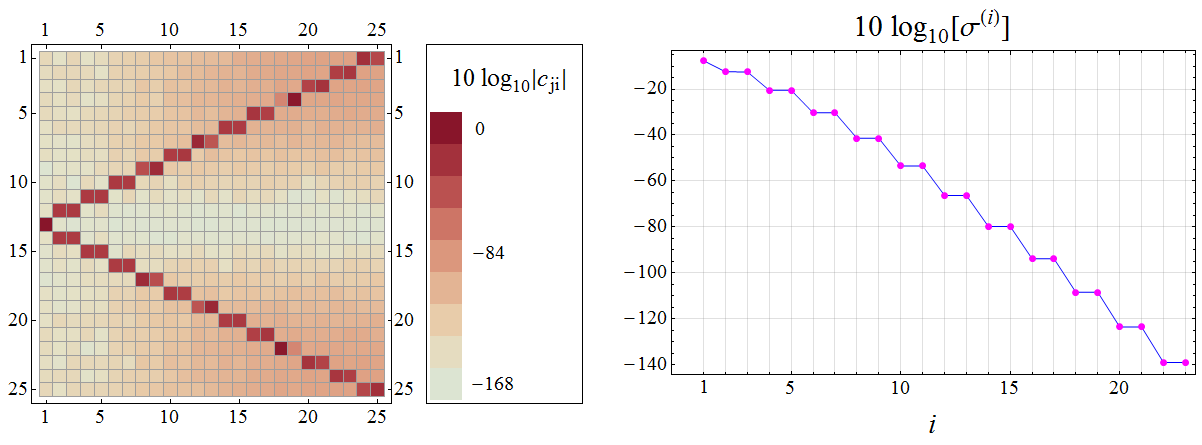}
\par\end{centering}
\caption{Spectral analysis of the channel operator (\ref{eq:channelMatrixCrossed}):
matrix plot of the scalar product $c_{ji}$ (left) and singular values
of $H_{pn}^{(\times)}$ as a function of the spectral index (right),
for $N=25$, $a=\lambda=1$ m, $l=\lambda/20$, $d=10\lambda$ and
$s_{R}=s_{T}^{*}=\mp i$.\label{fig:matrixPlotCrossed}}
\end{figure}

With the choice $s_{T}=\pm i$, $s_{R}=\mp i$, (\ref{eq:channelMatrixCrossed})
becomes:
\begin{eqnarray}
H_{pn}^{(\pm)} & = & \left[1+\frac{d^{2}}{d^{2}+2a^{2}\left\{ 1-\cos\left[\frac{2\pi\left(p-n\right)}{N}\right]\right\} }\right]\nonumber \\
 & \times & ik\eta\frac{I_{0}}{2}l^{2}\frac{\exp\left(-ik\sqrt{d^{2}+2a^{2}\left\{ 1-\cos\left[\frac{2\pi\left(p-n\right)}{N}\right]\right\} }\right)}{4\pi\sqrt{d^{2}+2a^{2}\left\{ 1-\cos\left[\frac{2\pi\left(p-n\right)}{N}\right]\right\} }},\label{eq:channelMatrixCrossedSimplified}
\end{eqnarray}
which is again a circulant matrix, as also evidenced by the strong
similarity between Figure \ref{fig:matrixPlotIsotropic} and Figure
\ref{fig:matrixPlotCrossed}.
\begin{figure}[!t]
\noindent \begin{centering}
\includegraphics[width=1\textwidth]{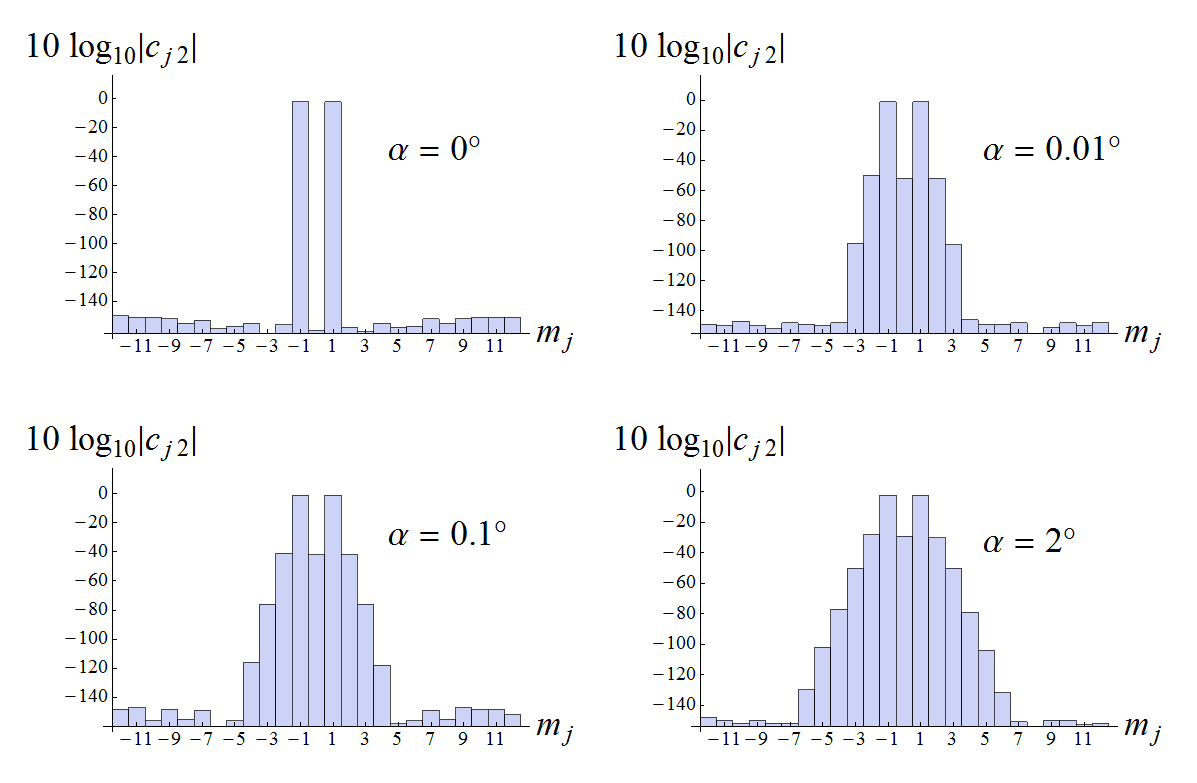}
\par\end{centering}
\caption{Histograms of the projection of the second right channel mode ($i=2$)
relative to a transmission link between two UCAs composed of $N=25$
crossed Hertzian dipoles over the vortex modes, for $a=\lambda=1$
m, $l=\lambda/20$, $d=10\lambda$, $s_{R}=s_{T}^{*}=-i$ and four
different rotation angles $\alpha$ of the receiving array around
the $y$-axis passing through its center. When $\alpha=0$ the two
UCAs are perfectly aligned.\label{fig:shiftComparison}}
\end{figure}
This fundamental result confirms that vortex modes are well-suited
to implement multiple communications between perfectly aligned UCAs
when circular polarization and low-directive elements are implied:
in this context, $\left\{ \left|m_{i}\right\rangle \right\} $ represents
a natural basis set of right channel modes yielding vortex waves with
good purity. Unfortunately, when expressed in terms of such basis,
the channel matrix spectrum is strongly influenced by the presence
of misalignments between the two arrays, as can be seen in Figure
\ref{fig:shiftComparison}.

\begin{figure}[!t]
\noindent \begin{centering}
\includegraphics[width=1\textwidth]{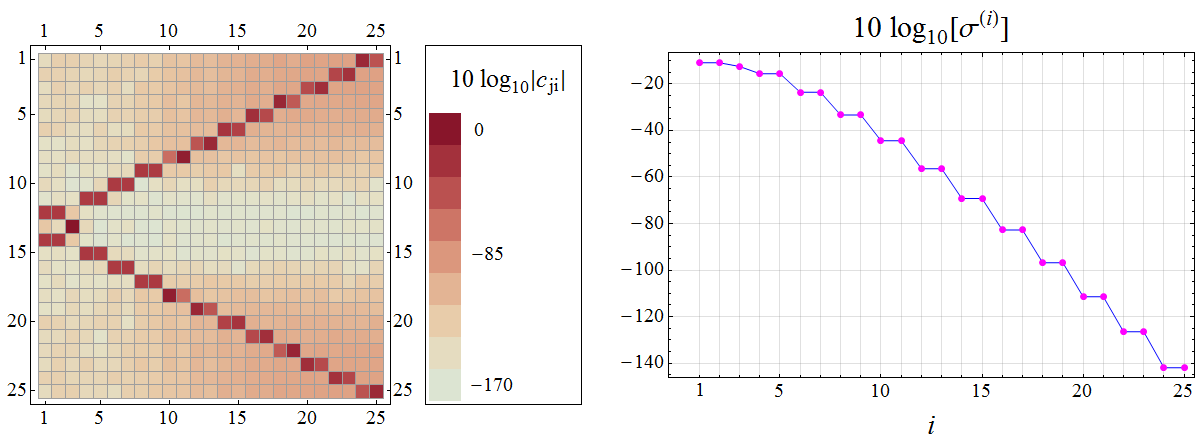}
\par\end{centering}
\caption{Spectral analysis of the channel operator (\ref{eq:matrixAzimuthally})
relative to a transmission link between perfectly aligned UCAs composed
of azimuthally polarized Hertzian dipoles: matrix plot of the scalar
product $c_{ji}$ (left) and singular values of $H_{pn}^{(A)}$ as
a function of the spectral index (right), for $N=25$, $a=\lambda=1$
m, $l=\lambda/20$ and $d=10\lambda$.\label{fig:matrixPlotAzi}}
\end{figure}
\begin{figure}[!t]
\noindent \begin{centering}
\includegraphics[width=1\textwidth]{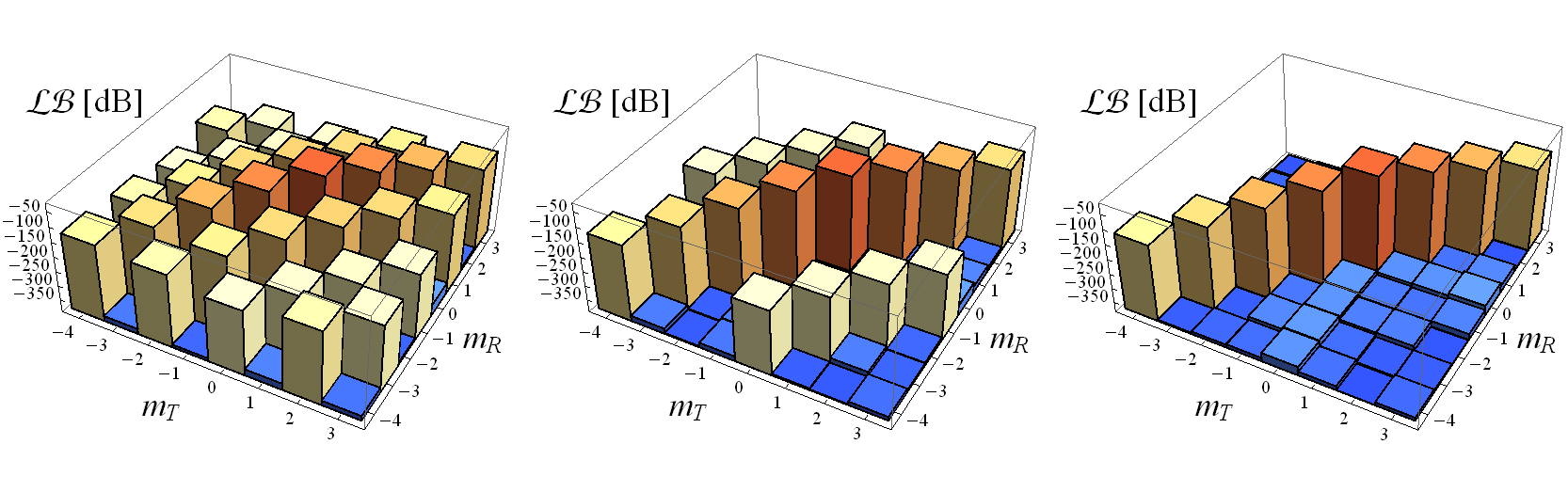}
\par\end{centering}
\caption{On-axis link budget relative to a communication link between two eight-element
UCAs as a function of the transmitting and receiving OAM excitation
coefficients for three different choices of the constituent antennas:
from left to right, $x$-directed half-wave dipoles, circularly polarized
crossed half-wave dipoles and circularly polarized crossed Hertzian
dipoles, respectively, for $\lambda=1.46$ m, $a=1.5$ m and $d=40$
m. \label{fig:linkBudgetComparison}}
\end{figure}

The use of circularly-polarized crossed Hertzian dipoles does not represent
the only way to enforce the symmetry condition so as to attain a pure
vortex spectrum of the channel matrix in communication links between
facing UCAs. For instance, azimuthally or radially polarized dipoles
are equally good candidates in this respect, the corresponding channel
matrices being expressed as in \ref{chap:AziRad}, formulas
(\ref{eq:matrixAzimuthally}) and (\ref{eq:matrixRadially}). The
projection of the right singular vectors relative to (\ref{eq:matrixAzimuthally})
on the vortex modes basis is reported in Figure \ref{fig:matrixPlotAzi}
(left) and differs from the previous ones in Figure \ref{fig:matrixPlotIsotropic}
and Figure \ref{fig:matrixPlotCrossed} owing to the atypical intensity
profile of the $m=0$ and $m=\pm1$ azimuthally polarized waves (compare
with Figure \ref{fig:aziRadUCAprofiles}). Analogous results are obtained
for the radially polarized case (\ref{eq:matrixRadially}).

Finally, in \ref{sec:Communication} it is explained how the knowledge
of the channel matrix and of the sets of transmitting and receiving
coefficients enables the estimation of the link budget, which represents
a useful means to check the mode orthogonality, as shown in Figure \ref{fig:linkBudgetComparison}.

\section{Conclusions}

Vortex waves are found to represent a natural set of solutions for
all the problems in which the circular symmetry is preserved, while
being very susceptible to departures from such condition. Contrary
to what is commonly believed, the use of UCAs does not represent a
sufficient constraint to generate vortex waves with good purity. The
presence of linear antennas is responsible for the symmetry
breaking of the vortex profiles and leads to a deterioration of the
on-axis vortex with charge $m$ in $m$ distinct off-axis vortices
with unitary charge. Owing to this symmetry breaking, the set of vortex
excitation coefficients (i.e., the columns of the DFT matrix) do no
longer represent a natural basis of channel modes for the communication
links between UCAs. A possible solution to restore the lost circular
symmetry requires the use of low-directive circularly, azimuthally or radially polarized sources. 

\ack
I would like to thank Dr. Rossella Gaffoglio, Prof. Francesco Andriulli and Prof. Giuseppe Vecchi, 
from the Polytechnic University of Torino, Dr. Assunta De Vita and Eng. Bruno Sacco, from the Centre for Research 
and Technological Innovation, RAI Radiotelevisione Italiana, for their precious teachings.

\appendix

\section{Green functions and the radiation problem\label{sec:GreenFunctions}}

In principle, when the source distribution is known, the inhomogeneous
wave equations can be solved via the Green function approach, enabling
the determination of the radiated fields \cite{Collin1969,Devaney2012,Jackson1999,Morse1953,Orfanidis2016}.
In the frequency domain, the outgoing-wave Green function is given
by:
\begin{equation}
g_{+}\left(\mathbf{R}\right)=-\frac{\exp\left(-ikR\right)}{4\pi R},\label{eq:GreenFunction}
\end{equation}
where $k=2\pi/\lambda$ is the wavenumber and $\lambda$ the wavelength.
The radiated scalar field can be formally written as:
\begin{equation}
f\left(\mathbf{r}\right)=f_{0}\left(\mathbf{r}\right)+\int d^{3}r'\:g_{+}\left(\mathbf{r}\text{\textminus}\mathbf{r}'\right)q\left(\mathbf{r}'\right),\label{eq:radiatedField}
\end{equation}
with $f_{0}\left(\mathbf{r}\right)$ representing an arbitrary
solution of the homogeneous Helmholtz equation and $q\left(\mathbf{r}\right)$
the source term. Similarly, an integral representation for the frequency-domain
electromagnetic potentials can be expressed in convolutional form:
\begin{equation}
\Phi\left(\mathbf{r}\right)=-\int d^{3}r'\:\frac{1}{\epsilon}\rho\left(\mathbf{r}'\right)g_{+}\left(\mathbf{r}\text{\textminus}\mathbf{r}'\right);
\end{equation}
\begin{equation}
\mathbf{A}\left(\mathbf{r}\right)=-\int d^{3}r'\:\mu\,\mathbf{J}\left(\mathbf{r}'\right)g_{+}\left(\mathbf{r}\text{\textminus}\mathbf{r}'\right),
\end{equation}
where $\rho$ and $\mathbf{J}$ stand for charge and current density
distributions and the principal value of the integrals is assumed.
The corresponding expressions for the electric and magnetic fields
in the Lorenz gauge read:
\begin{equation}
\mathbf{E}\left(\mathbf{r}\right)=\frac{i}{\omega\epsilon}\int d^{3}r'\:\left\{ \left[\mathbf{J}\left(\mathbf{r}'\right)\cdot\nabla\right]\nabla g_{+}\left(\mathbf{r}\text{\textminus}\mathbf{r}'\right)+k^{2}\mathbf{J}\left(\mathbf{r}'\right)g_{+}\left(\mathbf{r}\text{\textminus}\mathbf{r}'\right)\right\} ;\label{eq:electricGreenScalar}
\end{equation}
\begin{equation}
\mathbf{H}\left(\mathbf{r}\right)=\int d^{3}r'\:\mathbf{J}\left(\mathbf{r}'\right)\times\nabla g_{+}\left(\mathbf{r}\text{\textminus}\mathbf{r}'\right).\label{eq:magneticGreenScalar}
\end{equation}

Whenever it is necessary to determine the directional properties of
the radiated field, the following integral, known as the \textit{radiation
vector}, plays a leading role:
\begin{equation}
\mathbf{F}\left(\mathbf{k}\right)=\int d^{3}r'\:\mathbf{J}\left(\mathbf{r}'\right)\exp\left(i\mathbf{k}\cdot\mathbf{r}'\right).\label{eq:radiationVector}
\end{equation}
Indeed, it is possible to show that the polarization of the fields
radiated by an antenna depends on the transverse component of this
radiation vector, $\mathbf{F}_{\perp}\left(\mathbf{k}\right)$,
and a vector \textit{effective height} of the antenna can be defined
as: 
\begin{equation}
\mathbf{h}\left(\theta,\phi\right)=-\frac{1}{I_{0}}\mathbf{F}_{\perp}\left(\mathbf{k}\right)=-\frac{1}{I_{0}}\left(\mathbf{u}_{\theta}\mathbf{u}_{\theta}+\mathbf{u}_{\phi}\mathbf{u}_{\phi}\right)\cdot\mathbf{F}\left(\mathbf{k}\right),\label{eq:effectiveHeight}
\end{equation}
being $I_{0}$ the input current to the antenna terminals, $\theta$,
$\phi$, $\mathbf{u}_{\theta}$, $\mathbf{u}_{\phi}$ the
spherical angles and corresponding unit vectors and $\mathbf{u}\mathbf{u}$
the outer product $\mathbf{u}\otimes\mathbf{u}$. As an example,
for a $z$-directed half-wave dipole placed in the origin one gets:
\begin{equation}
\mathbf{h}\left(\theta,\phi\right)=\frac{2}{k\sin\theta}\cos\left(\frac{\pi}{2}\cos\theta\right)\mathbf{u}_{\theta},
\end{equation}
whereas for a square microstrip antenna laid in the $yz$ plane and
with feeding line parallel to the $y$-axis a little bit more tedious
calculation gives: 
\begin{equation}
\mathbf{h}\left(\theta,\phi\right)=-4L\cos\left(\frac{\pi L}{\lambda}\sin\theta\sin\phi\right)\mathrm{sinc}\left(\frac{\pi L}{\lambda}\cos\theta\right)\sin\theta\,\mathbf{u}_{\phi},\label{eq:microstrip}
\end{equation}
where $L$ represents the microstrip antenna side.

The evaluation of the fields radiated by a given source distribution
in free space and subject to the requirement of causality, often referred
to as ``the radiation problem'', can also be systematized through
an alternative approach involving the use of the dyadic Green functions \cite{Collin1969,Morse1953}.
For instance, if we define: 
\begin{equation}
\mathbf{g}_{+}^{E}\left(\mathbf{R}\right)=\left[\left(\mathbf{u}_{x}\mathbf{u}_{x}+\mathbf{u}_{y}\mathbf{u}_{y}+\mathbf{u}_{z}\mathbf{u}_{z}\right)+\frac{\nabla\nabla}{k^{2}}\right]\frac{\exp\left(-ikR\right)}{4\pi R},\label{eq:electricFieldGreenFunction}
\end{equation}
the electric field generated by an arbitrary current distribution
reads:
\begin{equation}
\mathbf{E}\left(\mathbf{r}\right)=-i\omega\mu\int d^{3}r'\:\mathbf{J}\left(\mathbf{r}'\right)\cdot\mathbf{g}_{+}^{E}\left(\mathbf{r}\text{\textminus}\mathbf{r}'\right),\label{eq:electricGreenDyadic}
\end{equation}
which is easily traced back to (\ref{eq:electricGreenScalar}). 
In both the Green function approaches, the source of the radiated
field is assumed to be specified and to be independent of the field
that it produces. However, a similar derivation can be obtained 
for both the electric and magnetic fields under the hypothesis 
of fictitious sources that appear as a result of the application 
of equivalence theorems. Within this framework, the radiation problem can be extended to
more complex scenarios, such as in the presence of physical obstacles,
dispersive media, anisotropic and inhomogeneous regions of space, 
by properly generalizing the above defined free-space
Green functions. 

\section{Communication by orthogonal modes\label{sec:Communication}}

Let us consider a set of $N$ distinct sources arranged over the region
$\mathcal{V}_{T}$ and $P$ receiving elements distributed within
$\mathcal{V}_{R}$. Suppose the sources and the receiving elements
to be characterized by vector effective heights $\mathbf{h}_{n}^{T}$
and $\mathbf{h}_{p}^{R}$, respectively, and spatial extent much
smaller than the distance $d$ between the two volumes. In correspondence
with a given set of excitation coefficients $\left\{ \xi_{n}\right\} \in\mathbb{C}^{N}$
for the transmitting array, the expression of the radiated electric
field is reported in (\ref{eq:Earray}). 

Assuming no coupling, the voltage induced on a $p$-th receiving element
in the absence of noise can be expressed approximately as the dot
product between the incoming electric field and the $p$-th effective
height \cite{Orfanidis2016}:
\begin{equation}
V_{p}=ik\eta I_{0}\sum_{n=1}^{N}\xi_{n}\frac{\exp\left(-ik\left|\mathbf{r}_{p}^{R}-\mathbf{r}_{n}^{T}\right|\right)}{4\pi\left|\mathbf{r}_{p}^{R}-\mathbf{r}_{n}^{T}\right|}\,\mathbf{h}_{n}^{T}\cdot\mathbf{h}_{p}^{R}.\label{eq:voltages}
\end{equation}
In (\ref{eq:voltages}), $\mathbf{r}_{p}^{R}$ identifies the
position of the $p$-th receiving element and the functions $\mathbf{h}_{n}^{T}$
and $\mathbf{h}_{p}^{R}$ are evaluated at the angular coordinates
which define the direction from the $n$-th source to the $p$-th
receiving element, and vice versa, according to the relative reference
frame. Equation (\ref{eq:voltages}) can be expressed in matrix form:
\begin{equation}
V_{p}=\sum_{n=1}^{N}H_{pn}\,\xi_{n},\label{eq:matrixEquation}
\end{equation}
with $H$ representing the so-called channel matrix for the considered
system:
\begin{equation}
H_{pn}=ik\eta I_{0}\,\frac{\exp\left(-ik\left|\mathbf{r}_{p}^{R}-\mathbf{r}_{n}^{T}\right|\right)}{4\pi\left|\mathbf{r}_{p}^{R}-\mathbf{r}_{n}^{T}\right|}\,\mathbf{h}_{n}^{T}\cdot\mathbf{h}_{p}^{R}.\label{eq:channelMatrix}
\end{equation}
The matrix elements $H_{pn}$ can of course be interpreted as coupling
coefficients between the $n$-th radiator and the $p$-th receiving
element, whose specific value depends on the type of antennas used
(i.e.\textit{,} the explicit form of the effective heights) and on
their geometrical arrangement and orientation within the two volumes
$\mathcal{V}_{T}$ and $\mathcal{V}_{R}$. 

If we suppose the transmitting array to be excited by means of an
ideal beamforming network, the total input power is given by:
\begin{equation}
\mathcal{P}_{in}=\frac{1}{2}RI_{0}^{2}\sum_{n=1}^{N}\left|\xi_{n}\right|^{2}=\frac{1}{2}RI_{0}^{2},\label{eq:Pin}
\end{equation}
being $R$ the resistance of the radiators and having required the
normalization condition for the synthesis coefficients. Similarly,
the total power collected by the receiving array under ideal circumstances
can be written as: 
\begin{equation}
\mathcal{P}_{out}=\frac{\left|V_{out}\right|^{2}}{8R}=\frac{1}{8R}\left|\sum_{p=1}^{P}\varUpsilon_{p}V_p\right|^{2}=\frac{1}{8R}\left|\sum_{p=1}^{P}\sum_{n=1}^{N}\varUpsilon_{p}\,H_{pn}\,\xi_{n}\right|^{2},\label{eq:Pout}
\end{equation}
where the elements of the set $\left\{ \varUpsilon_{p}\right\} \in\mathbb{C}^{P}$
are the beamforming weights introduced at the receiver and the network
is supposed to be connected to a matched load \cite{Orfanidis2016}.
The link budget of a communication system is defined as the ratio
between the received and the transmitted powers and plays a fundamental
role for the characterization of the link; in the present case, from
(\ref{eq:Pin}) and (\ref{eq:Pout}), the explicit expression of the
link budget reads: 
\begin{equation}
\mathcal{LB}=\frac{\mathcal{P}_{out}}{\mathcal{P}_{in}}=\left|\frac{1}{2RI_{0}}\sum_{p=1}^{P}\sum_{n=1}^{N}\varUpsilon_{p}\,H_{pn}\,\xi_{n}\right|^{2}.\label{eq:linkBudget}
\end{equation}

To find the transmitting and receiving column vectors $\left\{ \xi_{n}\right\} $
and $\left\{ \varUpsilon_{p}\right\} $ which define the sought channel
modes for the considered link between antenna arrays, it is possible
to resort to the singular value decomposition (SVD\nomenclature{SVD}{singular value decomposition})
of the channel matrix, thus rewriting it in terms of the product $H=U\Sigma W^{\dagger}$,
where $U$ and $W$ are unitary square matrices containing the left
and right singular vectors of $H$, respectively, and $\Sigma$ represents
a real diagonal rectangular matrix whose entries correspond to the
non-null singular values of $H$ sorted in decreasing order \cite{Golub1996,Horn1994}.
Such diagonal entries $\sigma^{(i)}=\Sigma_{ii}$ determine the gains
of the orthogonal subchannels identified by the couples $\left\{ \xi_{n}^{(i)},\varUpsilon_{p}^{(i)}\right\} =\left\{ W_{ni},U_{pi}^{*}\right\} $. 

It is straightforward to verify that, when the transmitting and receiving
beamforming weights are chosen to be the so-derived channel modes,
the link budget formula (\ref{eq:linkBudget}) for a communication
by the $i$-th subchannel reduces to:
\begin{equation}
\mathcal{LB}^{(i)}=\left(\frac{\sigma^{(i)}}{2RI_{0}}\right)^{2}.\label{eq:spectralLinkBudget}
\end{equation}
Equation (\ref{eq:spectralLinkBudget}) is just the proof that channel
modes represent the best coupled pairs of transmitting and receiving
beamforming coefficients, since the link budgets for the corresponding
communications are the optimal ones, as follows from the SVD definition. 

A sum rule is easily derived by considering the superpositions of
all the possible link budgets from the $n$-th source to the $p$-th
receiver, obtained from (\ref{eq:linkBudget}) with $\xi$ and $\varUpsilon$
replaced by a vector of $N-1$ zeros and a one at position $n$ and
by a vector of $P-1$ zeros and a one at position $p$, respectively:
\begin{equation}
\sum_{p=1}^{P}\sum_{n=1}^{N}\mathcal{LB}^{(n\rightarrow p)}=\sum_{p=1}^{P}\sum_{n=1}^{N}\left|\frac{H_{pn}}{2RI_{0}}\right|^{2}=\mathrm{tr}\left[\frac{H^{\dagger}H}{\left(2RI_{0}\right)^{2}}\right]=\sum_{i}\left(\frac{\sigma^{(i)}}{2RI_{0}}\right)^{2}=\sum_{i}\mathcal{LB}^{(i)}.\label{eq:matrixSumRule}
\end{equation}


The voltage induced on the receiving antenna by the incoming wave has 
been expressed in (\ref{eq:voltages}) as the
dot product between the unperturbed electric field 
and the antenna effective height. This is of course an approximation,
the scattered field being neglected under the assumption
of a single plane wave incidence. Although such approximation proves
to be quite convenient to deal with antenna arrays, it is reasonable
to expect that it gives accurate results to the extent that the receiving
antennas do not alter consistently the background electric field and
can be considered almost point-like, two conditions which are hardly
met in a realistic situation. The link budget concept can be extended 
beyond this simplified scenario with the help of some fundamental results
of the electromagnetic theory, enabling a rigorous definition which
applies, in principle, to the majority of communication systems, including
those based on aperture or reflector antennas, spiral phase plates,
lenses and analogous optical devices. 

Let us consider an arbitrary receiving antenna together with its surrounding
structure and a background electromagnetic field $\left\{ \mathbf{E}_{inc}\left(\mathbf{r}\right),\mathbf{H}_{inc}\left(\mathbf{r}\right)\right\} $ impinging on it. A more precise formulation of the received voltage is given by:
\begin{equation}
V_{out}=-\int d\mathbf{l}\cdot\mathbf{E}_{a}\left(\mathbf{l}\right),\label{eq:TheveninVoltage}
\end{equation}
where $\mathbf{E}_{a}$ represents the electric field in the presence
of the open circuited antenna and its surroundings and the integral is performed over a line
connecting the antenna terminals. Making use of the well-known electromagnetic
equivalence principles \cite{Collin1969}, a generalization of equation
(\ref{eq:electricGreenDyadic}) is now provided by expressing the
total electric field inside an arbitrary finite volume $\mathcal{V}$
which contains the antenna and any material objects in its proximity
as if it were generated by electric $\mathbf{j}_{\mathcal{S}}$, 
in units of A/m, and magnetic $\mathbf{m}_{\mathcal{S}}$, in V/m, equivalent surface currents
distributed over the volume boundary $\mathcal{S}=\partial\mathcal{V}$: 
\begin{equation}
\mathbf{E}_{a}\left(\mathbf{r}_{\mathcal{V}}\right)=\int_{\mathcal{S}}d\mathbf{r}_{\mathcal{S}}\left[\mathbf{j}_{\mathcal{S}}\left(\mathbf{r}_{\mathcal{S}}\right)\cdot\mathbf{g}_{j}^{E_{a}}\left(\mathbf{r}_{\mathcal{V}},\mathbf{r}_{\mathcal{S}}\right)+\mathbf{m}_{\mathcal{S}}\left(\mathbf{r}_{\mathcal{S}}\right)\cdot\mathbf{g}_{m}^{E_{a}}\left(\mathbf{r}_{\mathcal{V}},\mathbf{r}_{\mathcal{S}}\right)\right].\label{eq:totalEfield}
\end{equation}
In (\ref{eq:totalEfield}), $\mathbf{g}_{j}^{E_{a}}$ and $\mathbf{g}_{m}^{E_{a}}$
represent the electric field dyadic Green functions relative to the
radiation from the equivalent electric and magnetic sources in the
presence of the antenna and its surroundings, $\mathbf{r}_{\mathcal{V}}$
is a generic point inside $\mathcal{V}$ and $\mathbf{r}_{\mathcal{S}}$
the integration variable over $\mathcal{S}$. Furthermore, all the
dimensional constants are now incorporated in the Green functions
definitions, in contrast to the convention employed for the free-space
equation (\ref{eq:electricGreenDyadic}). 

The use of equivalent surface currents permits to postulate the existence
of a different electromagnetic field $\left\{ \mathbf{E}_{ext},\mathbf{H}_{ext}\right\} $
outside $\mathcal{V}$.
In other words, the original problem, where the electromagnetic field
$\left\{ \mathbf{E}_{inc},\mathbf{H}_{inc}\right\} $ generated
by some distant sources scatters the antenna and its structure giving
rise to a total resultant field $\left\{ \mathbf{E}_{a},\mathbf{H}_{a}\right\} $,
has been replaced by an equivalent one in which the fictitious sources
$\mathbf{j}_{\mathcal{S}}$ and $\mathbf{m}_{\mathcal{S}}$
over $\mathcal{S}$ radiate $\left\{ \mathbf{E}_{a},\mathbf{H}_{a}\right\} $
inside $\mathcal{V}$ (in the presence of the antenna and its surroundings)
and $\left\{ \mathbf{E}_{ext},\mathbf{H}_{ext}\right\} $ in the
free space outside $\mathcal{V}$. In order that the so defined field
be a valid solution to the Maxwell equations throughout the whole
space, the equivalent sources on the volume boundary must ensure the
discontinuous change in the tangential components of the electric
and magnetic fields across $\mathcal{S}$: 
\begin{equation}
\mathbf{j}_{\mathcal{S}}=\mathbf{u}_{n}\times\left(\mathbf{H}_{ext}-\mathbf{H}_{a}\right);\quad\mathbf{m}_{\mathcal{S}}=-\mathbf{u}_{n}\times\left(\mathbf{E}_{ext}-\mathbf{E}_{a}\right),
\end{equation}
where $\mathbf{u}_{n}$ is the unit vector associated with the
outward pointing normal to the surface $\mathcal{S}$. 

By choosing $\left\{ \mathbf{E}_{ext},\mathbf{H}_{ext}\right\} $
to be the scattered electromagnetic field $\left\{ \mathbf{E}_{a}-\mathbf{E}_{inc},\mathbf{H}_{a}-\mathbf{H}_{inc}\right\}$,
the equivalent sources are then easily expressed via the incident electromagnetic
field in the absence of the antenna and its structure:
\begin{equation}
\mathbf{j}_{\mathcal{S}}=-\mathbf{u}_{n}\times\mathbf{H}_{inc};\quad\mathbf{m}_{\mathcal{S}}=\mathbf{u}_{n}\times\mathbf{E}_{inc}.\label{eq:equivalentSources}
\end{equation}
Several approaches can be adopted to express the received voltage (\ref{eq:TheveninVoltage}) as a reaction integral \cite{Harrington2001}; among those, the following one is of particular interest:
\begin{equation}
V_{out}=\frac{1}{I_{0}}\int_{\mathcal{S}}d\mathbf{r}_{\mathcal{S}}\left[\mathbf{H}_{a}^{0}\left(\mathbf{r}_{\mathcal{S}}\right)\cdot\mathbf{m}_{\mathcal{S}}\left(\mathbf{r}_{\mathcal{S}}\right)-\mathbf{E}_{a}^{0}\left(\mathbf{r}_{\mathcal{S}}\right)\cdot\mathbf{j}_{\mathcal{S}}\left(\mathbf{r}_{\mathcal{S}}\right)\right],\label{eq:reactionIntegral}
\end{equation}
since it only involves (\ref{eq:equivalentSources}) and the electromagnetic
field $\left\{ \mathbf{E}_{a}^{0},\mathbf{H}_{a}^{0}\right\} $
radiated by an impressed electric current term $\mathbf{j}_{0}$
with amplitude $I_{0}$ on the antenna terminals in the presence of
both the antenna and its surroundings. More explicitly, the integrand
in (\ref{eq:reactionIntegral}) has the form of a dot product between
the field radiated by the receiving antenna in transmission and the
tangential components of the incoming field across the surface $\mathcal{S}$:
\begin{equation}
V_{out}=\frac{1}{I_{0}}\int_{\mathcal{S}}d\mathbf{r}_{\mathcal{S}}\left[\mathbf{H}_{a}^{0}\cdot\left(\mathbf{u}_{n}\times\mathbf{E}_{inc}\right)+\mathbf{E}_{a}^{0}\cdot\left(\mathbf{u}_{n}\times\mathbf{H}_{inc}\right)\right].\label{eq:reactionIntegralExplicit}
\end{equation}
The proof of equation (\ref{eq:reactionIntegral}) is a straightforward
application of the reciprocity theorem for the pairs of sources $\left\{ \mathbf{j}_{\mathcal{S}},\mathbf{m}_{\mathcal{S}}\right\} $,
$\left\{ \mathbf{j}_{0}\right\} $ and fields $\left\{ \mathbf{E}_{inc},\mathbf{H}_{inc}\right\} $,
$\left\{ \mathbf{E}_{a}^{0},\mathbf{H}_{a}^{0}\right\} $.
On using some basic vector formulas, the above expression can be rewritten
in terms of a flux integral: 
\begin{equation}
V_{out}=\frac{1}{I_{0}}\int_{\mathcal{S}}d\mathbf{r}_{\mathcal{S}}\left(\mathbf{E}_{inc}\times\mathbf{H}_{a}^{0}-\mathbf{E}_{a}^{0}\times\mathbf{H}_{inc}\right)\cdot\mathbf{u}_{n}.\label{eq:reactionIntegralEH}
\end{equation}
The total received power is derived from (\ref{eq:reactionIntegralEH})
by means of the usual formula:
\begin{equation}
\mathcal{P}_{out}=\frac{\left|V_{out}\right|^{2}}{8R}=\frac{\left|\int_{\mathcal{S}}d\mathbf{r}_{\mathcal{S}}\left(\mathbf{E}_{inc}\times\mathbf{H}_{a}^{0}-\mathbf{E}_{a}^{0}\times\mathbf{H}_{inc}\right)\cdot\mathbf{u}_{n}\right|^{2}}{4\int_{\mathcal{S}}d\mathbf{r}_{\mathcal{S}}\left(\mathbf{E}_{a}^{0}\times\mathbf{H}_{a}^{0*}+\mathbf{E}_{a}^{0*}\times\mathbf{H}_{a}^{0}\right)\cdot\mathbf{u}_{n}},\label{eq:generalizedPout}
\end{equation}
where $I_{0}$ has been expressed through the Poynting vector flux
under the hypothesis of a perfectly efficient antenna: 
\begin{equation}
\mathcal{P}_{0}=\frac{1}{2}RI_{0}^{2}=\frac{1}{2}\int_{\mathcal{S}}d\mathbf{r}_{\mathcal{S}}\left[\mathrm{Re}\left(\mathbf{E}_{a}^{0}\times\mathbf{H}_{a}^{0*}\right)\right]\cdot\mathbf{u}_{n}.\label{eq:Poynting}
\end{equation}
An extended link budget formula is then simply obtained by performing
the ratio between (\ref{eq:generalizedPout}) and the total power
delivered to the source, which in turn can be described as a power
outflow at the transmitting side. The importance of equation (\ref{eq:generalizedPout})
lies in the fact that it is completely general and can be computed
from the only knowledge of the incident electromagnetic field and
of the field radiated by the receiving antenna when operating in transmission
mode. 


\section{Azimuthally and radially polarized dipoles\label{chap:AziRad}}

In Section \ref{sec:channelMatrices}, some channel matrices relative
to a communication link between facing UCAs have been presented for
different choices in the disposition of the arrays elements. When
azimuthally and radially polarized Hertzian dipoles are involved,
the corresponding channel matrices read:

\begin{equation}
H_{pn}^{(A)}=H_{pn}^{(\circ)}\sin\theta_{np}^{\varphi}\sin\theta_{pn}^{\varphi}\,\boldsymbol{\theta}_{np}^{\varphi}\cdot\boldsymbol{\theta}_{pn}^{\varphi};\label{eq:matrixAzimuthally}
\end{equation}
\begin{equation}
H_{pn}^{(R)}=H_{pn}^{(\circ)}\sin\theta_{np}^{\rho}\sin\theta_{pn}^{\rho}\,\boldsymbol{\theta}_{np}^{\rho}\cdot\boldsymbol{\theta}_{pn}^{\rho},\label{eq:matrixRadially}
\end{equation}
where $H_{pn}^{(\circ)}$ is defined in (\ref{eq:channelMatrixIsotropic})
and we now have:
\begin{eqnarray}
\mathbf{r}_{np} & \!\!=\!\! & \left\{ x_{np},y_{np},z_{np}\right\} =\mathbf{r}_{p}^{R}-\mathbf{r}_{n}^{T};\\
\mathbf{v}_{n}^{\varphi} & \!\!=\!\! & \left\{ \mathbf{x}_{n}^{\varphi},\mathbf{y}_{n}^{\varphi},\mathbf{z}_{n}^{\varphi}\right\} =\mathsf{R}_{z}^{-1}\left(\varphi_{n}+\frac{\pi}{2}\right)\left\{ \mathbf{u}_{x},\mathbf{u}_{y},\mathbf{u}_{z}\right\} ;\\
\theta_{np}^{\varphi} & \!\!=\!\! & \arccos\left(\mathbf{x}_{n}^{\varphi}\cdot\mathbf{r}_{np}/r_{np}\right);\\
\phi_{np}^{\varphi} & \!\!=\!\! & \arctan\left(\mathbf{z}_{n}^{\varphi}\cdot\mathbf{r}_{np}/\mathbf{y}_{n}^{\varphi}\cdot\mathbf{r}_{np}\right);\\
\boldsymbol{\theta}_{np}^{\varphi} & \!\!=\!\! & \mathsf{R}_{z}\left(\varphi_{n}+\frac{\pi}{2}\right)\left\{ -\sin\theta_{np}^{\varphi},\cos\theta_{np}^{\varphi}\cos\phi_{np}^{\varphi},\cos\theta_{np}^{\varphi}\sin\phi_{np}^{\varphi}\right\} ;
\end{eqnarray}
\begin{eqnarray}
\mathbf{v}_{p}^{\varphi} & \!\!=\!\! & \left\{ \mathbf{x}_{p}^{\varphi},\mathbf{y}_{p}^{\varphi},\mathbf{z}_{p}^{\varphi}\right\} =\mathsf{R}_{z}^{-1}\left(\varphi_{p}+\frac{\pi}{2}\right)\left\{ \mathbf{u}_{x},\mathbf{u}_{y},\mathbf{u}_{z}\right\} ;\\
\theta_{pn}^{\varphi} & \!\!=\!\! & \arccos\left(\mathbf{x}_{p}^{\varphi}\cdot\mathbf{r}_{pn}/r_{pn}\right);\\
\phi_{pn}^{\varphi} & \!\!=\!\! & \arctan\left(\mathbf{z}_{p}^{\varphi}\cdot\mathbf{r}_{pn}/\mathbf{y}_{p}^{\varphi}\cdot\mathbf{r}_{pn}\right);\\
\boldsymbol{\theta}_{pn}^{\varphi} & \!\!=\!\! & \mathsf{R}_{z}\left(\varphi_{p}+\frac{\pi}{2}\right)\left\{ -\sin\theta_{pn}^{\varphi},\cos\theta_{pn}^{\varphi}\cos\phi_{pn}^{\varphi},\cos\theta_{pn}^{\varphi}\sin\phi_{pn}^{\varphi}\right\} ;
\end{eqnarray}
\begin{eqnarray}
\mathbf{v}_{n}^{\rho} & \!\!=\!\! & \left\{ \mathbf{x}_{n}^{\rho},\mathbf{y}_{n}^{\rho},\mathbf{z}_{n}^{\rho}\right\} =\mathsf{R}_{z}^{-1}\left(\varphi_{n}\right)\left\{ \mathbf{u}_{x},\mathbf{u}_{y},\mathbf{u}_{z}\right\} ;\\
\theta_{np}^{\rho} & \!\!=\!\! & \arccos\left(\mathbf{x}_{n}^{\rho}\cdot\mathbf{r}_{np}/r_{np}\right);\\
\phi_{np}^{\rho} & \!\!=\!\! & \arctan\left(\mathbf{z}_{n}^{\rho}\cdot\mathbf{r}_{np}/\mathbf{y}_{n}^{\rho}\cdot\mathbf{r}_{np}\right);\\
\boldsymbol{\theta}_{np}^{\rho} & \!\!=\!\! & \mathsf{R}_{z}\left(\varphi_{n}\right)\left\{ -\sin\theta_{np}^{\rho},\cos\theta_{np}^{\rho}\cos\phi_{np}^{\rho},\cos\theta_{np}^{\rho}\sin\phi_{np}^{\rho}\right\} ;
\end{eqnarray}
\begin{eqnarray}
\mathbf{v}_{p}^{\rho} & \!\!=\!\! & \left\{ \mathbf{x}_{p}^{\rho},\mathbf{y}_{p}^{\rho},\mathbf{z}_{p}^{\rho}\right\} =\mathsf{R}_{z}^{-1}\left(\varphi_{p}\right)\left\{ \mathbf{u}_{x},\mathbf{u}_{y},\mathbf{u}_{z}\right\} ;\\
\theta_{pn}^{\rho} & \!\!=\!\! & \arccos\left(\mathbf{x}_{p}^{\rho}\cdot\mathbf{r}_{pn}/r_{pn}\right);\\
\phi_{pn}^{\rho} & \!\!=\!\! & \arctan\left(\mathbf{z}_{p}^{\rho}\cdot\mathbf{r}_{pn}/\mathbf{y}_{p}^{\rho}\cdot\mathbf{r}_{pn}\right);\\
\boldsymbol{\theta}_{pn}^{\rho} & \!\!=\!\! & \mathsf{R}_{z}\left(\varphi_{p}\right)\left\{ -\sin\theta_{pn}^{\rho},\cos\theta_{pn}^{\rho}\cos\phi_{pn}^{\rho},\cos\theta_{pn}^{\rho}\sin\phi_{pn}^{\rho}\right\} ,
\end{eqnarray}
with $\mathsf{R}_{z}\left(\varphi_{n}\right)$ representing the rotation
matrix by an angle $2\pi\left(n-1\right)/N$ around the $z$-axis.
After some lengthy calculations, we get: 
\begin{equation}
H_{pn}^{(A)}=H_{pn}^{(\circ)}\,\frac{a^{2}\left\{ 3+\cos\left[\frac{4\pi\left(p-n\right)}{N}\right]\right\} -2\left(2a^{2}+d^{2}\right)\cos\left[\frac{2\pi\left(p-n\right)}{N}\right]}{4a^{2}\cos\left[\frac{2\pi\left(p-n\right)}{N}\right]-2\left(2a^{2}+d^{2}\right)};\label{eq:matAziSimplified}
\end{equation}
\begin{equation}
H_{pn}^{(R)}=H_{pn}^{(\circ)}\,\frac{d^{2}\cos\left[\frac{2\pi\left(p-n\right)}{N}\right]+a^{2}\sin^{2}\left[\frac{2\pi\left(p-n\right)}{N}\right]}{d^{2}+2a^{2}\left\{ 1-\cos\left[\frac{2\pi\left(p-n\right)}{N}\right]\right\} }.\label{eq:matRadSimplified}
\end{equation}
Since both $H_{pn}^{(A)}$ and $H_{pn}^{(R)}$ depend on their indices
only through the difference $\left(p-n\right)$, they are circulant
and can be diagonalized via DFT matrix. 

It is worth mentioning that a possible alternative to the above approach
consists in the use of properly fed crossed dipoles, in which case:
\begin{equation}
H_{pn}^{(A)}=\left(-\sin\varphi_{n}\mathbf{E}_{n}^{x}+\cos\varphi_{n}\mathbf{E}_{n}^{y}\right)\cdot\left(-\sin\varphi_{p}\mathbf{h}_{p}^{x}+\cos\varphi_{p}\mathbf{h}_{p}^{y}\right);\label{eq:matAziCros}
\end{equation}
\begin{equation}
H_{pn}^{(R)}=\left(\cos\varphi_{n}\mathbf{E}_{n}^{x}+\sin\varphi_{n}\mathbf{E}_{n}^{y}\right)\cdot\left(\cos\varphi_{p}\mathbf{h}_{p}^{x}+\sin\varphi_{p}\mathbf{h}_{p}^{y}\right),\label{eq:matRadCros}
\end{equation}
with the same convention as in (\ref{eq:crossedDipolesMatrix}). When
explicitly rearranged, equations (\ref{eq:matAziCros}) and (\ref{eq:matRadCros})
still lead to (\ref{eq:matAziSimplified}) and (\ref{eq:matRadSimplified}),
respectively.

\section{Channel matrix degeneracy\label{sec:degeneracy}}

When the number of transmitting and receiving antennas is the same,
a diagonalization procedure can be considered instead of the full
SVD (whenever possible), in which case both the right and left singular
vectors are reinterpreted in terms of the matrix eigenvectors. An
instructive and straightforward example in this sense is provided
by the channel matrix (\ref{eq:channelMatrixCrossedSimplified}):
it can be shown that $H_{pn}^{(\pm)}$ is normal, which means that
a rigorous connection exists between its diagonalization and the SVD
\cite{Horn1994}. As mentioned above, $H_{pn}^{(\pm)}$ is diagonalized
by the DFT matrix $T_{pi}=\varPhi_{m_{i}}^{p}$ and, therefore, equation
(\ref{eq:linkBudget}) can be rewritten as: 
\begin{equation}
\mathcal{LB}=\left|\frac{1}{2RI_{0}}\sum_{p,i,j,n=1}^{N}\varUpsilon_{p}\,T_{pi}\,D_{ij}\,T_{jn}^{\dagger}\,\xi_{n}\right|^{2},\label{eq:LBcalculation}
\end{equation}
where $D_{ij}$ corresponds to the diagonal matrix containing the
eigenvalues of $H_{pn}^{(\pm)}$. Since the $N$ columns of the DFT
matrix are nothing but the vectors $\left|m_{i}\right\rangle $ defined
according to (\ref{eq:mConvention}), the best pair of transmitting
and receiving coefficients is given by the conjugate couple of vortex
excitations: 
\begin{equation}
\left\{ \xi_{n}^{m_{i}},\varUpsilon_{p}^{m_{i}}\right\} =\left\{ T_{ni},T_{ip}^{\dagger}\right\} =\left\{ \varPhi_{m_{i}}^{n},\left(\varPhi_{m_{i}}^{p}\right)^{*}\right\} =\left\{ \varPhi_{m_{i}}^{n},\varPhi_{-m_{i}}^{p}\right\} ,\label{eq:conjugateCouple}
\end{equation}
a result which is intuitively consistent with the request for a mode-matched
reception. Substituting expression (\ref{eq:conjugateCouple})
into (\ref{eq:LBcalculation}) and making use of the Dirac notation,
we get: 
\begin{eqnarray}
\mathcal{LB}^{(m_{i})} & = & \left|\frac{\left\langle m_{i}\right|H^{(\pm)}\left|m_{i}\right\rangle }{2RI_{0}}\right|^{2}=\left|\frac{1}{2RI_{0}}\left\langle m_{i}\right|\left(\sum_{j}\lambda^{(j)}\left|m_{j}\right\rangle \left\langle m_{j}\right|\right)\left|m_{i}\right\rangle \right|^{2}\nonumber \\
 & = & \left|\frac{1}{2RI_{0}}\left(\sum_{j}\lambda^{(j)}\left\langle m_{i}\right.\left|m_{j}\right\rangle \left\langle m_{j}\right.\left|m_{i}\right\rangle \right)\right|^{2}=\left(\frac{\left|\lambda^{(i)}\right|}{2RI_{0}}\right)^{2},\label{eq:spectralLinkBudgetOAM}
\end{eqnarray}
where $\lambda^{(i)}$ represents the $i$-th eigenvalue of the channel
matrix and the orthogonality relation in (\ref{eq:DiracOrthogonalityCompleteness})
has been exploited. Upon proper rearrangement of the eigenvalues list,
(\ref{eq:spectralLinkBudgetOAM}) is still the same as (\ref{eq:spectralLinkBudget}),
being the singular values of a normal matrix equal to the modulus
of its eigenvalues \cite{Horn1994}. 
\begin{figure}[!t]
\noindent \begin{centering}
\includegraphics[width=1\textwidth]{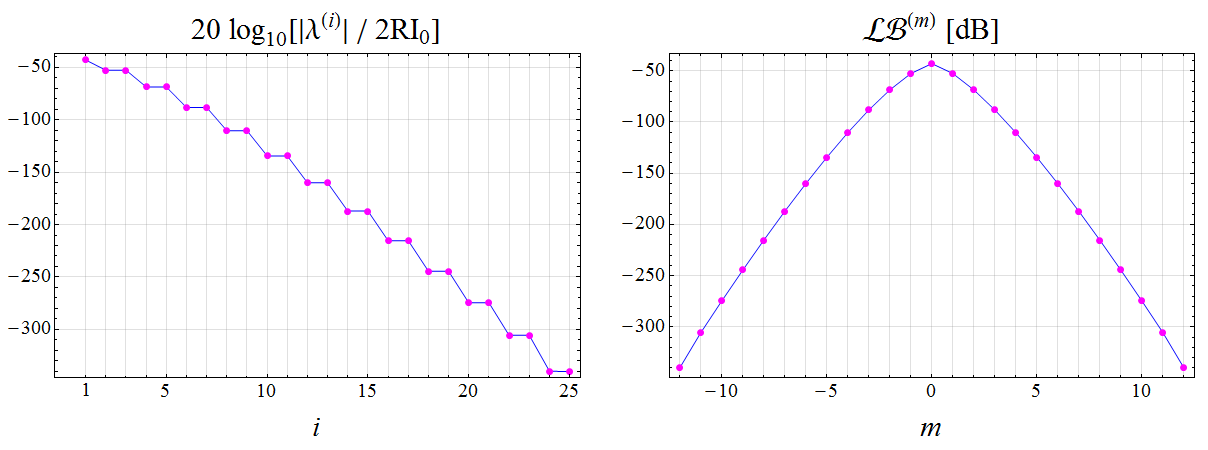}
\par\end{centering}
\caption{Comparison between the eigenvalues of (\ref{eq:channelMatrixCrossedSimplified}),
expressed in terms of the spectral index $i$ (left), and the link budget
(\ref{eq:linkBudget}) relative to $H_{pn}^{(\pm)}$ for the couple
of transmitting and receiving beamforming coefficients defined in
(\ref{eq:conjugateCouple}) as a function of the OAM charge $m$ (right),
for $N=25$, $a=\lambda=1$ m, $l=\lambda/20$, $d=10\lambda$.\label{fig:linkEigenComparison}}
\end{figure}

In Figure \ref{fig:linkEigenComparison}, a comparison between the
eigenvalues distribution of $H_{pn}^{(\pm)}$ and the corresponding
link budget for the communication by vortex waves as a function of
$m$ is proposed for the same choice of the parameters introduced
in the main text. From the comparison we infer how the two-fold degeneracy
in the channel matrix spectrum can be traced back to the fact that
two communications by vortex waves with opposite charges $\pm m$
share the same power level. 

Equation (\ref{eq:conjugateCouple}) tells us that, at least when
identical facing UCAs are concerned, the best beamforming weights
that should be introduced at the receiver to properly collect an incoming
vortex wave are just the complex conjugates of those employed for
generating such wave at the transmitting side\footnote{The complex conjugation results from the chirality inversion upon
reception together with the circumstance that the same orientation
direction has been chosen in the labeling of the elements along both
the transmitting and the receiving UCAs. A different approach would
be to employ an opposite sorting for the two arrays, in which case
the complex conjugation is no longer present.}. The validity of this result extends far beyond the proposed scenario,
in that it represents a natural consequence of the
reciprocity theorem.\\\\


\begin{thebibliography}{10}
\expandafter\ifx\csname url\endcsname\relax
  \def\url#1{{\tt #1}}\fi
\expandafter\ifx\csname urlprefix\endcsname\relax\def\urlprefix{URL }\fi
\providecommand{\eprint}[2][]{\url{#2}}

\bibitem{Allen2003}
Allen L, Barnett S~M and Padgett M~J 2003 {\em Optical Angular Momentum\/}
  (Institute of Physics Publishing, Bristol)

\bibitem{Andrews2008}
Andrews D~L 2008 {\em Structured Light and Its Applications: An Introduction to
  Phase-Structured Beams and Nanoscale Optical Forces\/} (Academic Press,
  Elsevier, USA)

\bibitem{Andrews2013}
Andrews D~L and Babiker M 2013 {\em The Angular Momentum of Light\/} (Cambridge
  University Press)

\bibitem{Torres2011}
Torres J~P and Torner L 2011 {\em Twisted Photons\/} (WILEY-VCH Verlag and Co.
  KGaA, Weinheim, Germany)

\bibitem{Bazhenov1990}
Bazhenov V~Y, Vasnetsov M~V and Soskin M~S 1990 {\em Journal of Experimental
  and Theoretical Physics Letters\/} {\bf 52} 429--431

\bibitem{Beijersbergen1993}
Beijersbergen M~W, Allen L, van~der Veen H~E~L~O and Woerdman J~P 1993 {\em
  Optics Communications\/} {\bf 96} 123--132
  \urlprefix\url{http://www.sciencedirect.com/science/article/pii/003040189390535D}

\bibitem{Beijersbergen1994}
Beijersbergen M~W, Coerwinkel R~P~C, Kristensen M and Woerdman J~P 1994 {\em
  Optics Communications\/} {\bf 112} 321--327
  \urlprefix\url{http://www.sciencedirect.com/science/article/pii/0030401894906386}

\bibitem{Padgett2000}
Padgett M and Allen L 2000 {\em Contemporary Physics\/} {\bf 41} 275--285
  \urlprefix\url{https://doi.org/10.1080/001075100750012777}

\bibitem{Yao2011}
Yao A~M and Padgett M~J 2011 {\em Adv. Opt. Photon.\/} {\bf 3} 161--204
  \urlprefix\url{http://aop.osa.org/abstract.cfm?URI=aop-3-2-161}

\bibitem{Marrucci2011}
Marrucci L, Karimi E, Slussarenko S, Piccirillo B, Santamato E, Nagali E and
  Sciarrino F 2011 {\em Journal of Optics\/} {\bf 13}
  \urlprefix\url{http://stacks.iop.org/2040-8986/13/i=6/a=064001}

\bibitem{Marrucci2006}
Marrucci L, Manzo C and Paparo D 2006 {\em Physical Review Letters\/} {\bf
  96}(16)
  \urlprefix\url{https://link.aps.org/doi/10.1103/PhysRevLett.96.163905}

\bibitem{Berry1987}
Berry M~V 1987 {\em Journal of Modern Optics\/} {\bf 34} 1401--1407
  \urlprefix\url{https://doi.org/10.1080/09500348714551321}

\bibitem{Barbuto2014}
Barbuto M, Trotta F, Bilotti F and Toscano A 2014 {\em Progress in
  Electromagnetics Research\/} {\bf 148} 23--30
  \urlprefix\url{http://www.jpier.org/PIER/pier.php?paper=14050204}

\bibitem{Cheng2014}
Cheng L, Hong W and Hao Z~C 2014 {\em Scientific Reports\/} {\bf 4}
  \urlprefix\url{https://www.nature.com/articles/srep04814}

\bibitem{Fonseca2015}
Fonseca N~J~G, Coulomb L and Angevain J~C 2015 A {Fresnel-like} reflector
  antenna design for high-order orbital angular momentum states {\em EuCAP 2015
  - The 9th European Conference on Antennas and Propagation\/}
  \urlprefix\url{https://ieeexplore.ieee.org/abstract/document/7228413/}

\bibitem{Maccalli2013}
Maccalli S, Pisano G, Colafrancesco S, Maffei B, Richard~Ng M~W and Gray M 2013
  {\em Applied Optics\/} {\bf 52} 635--639
  \urlprefix\url{http://ao.osa.org/abstract.cfm?URI=ao-52-4-635}

\bibitem{Niemiec2014}
Niemiec R, Brousseau C, Mahdjoubi K, Emile O and M\'{e}nard A 2014 {\em IEEE
  Antennas and Wireless Propagation Letters\/} {\bf 13} 1011--1014

\bibitem{Tennant2012}
Tennant A and Allen B 2012 {\em Electronics Letters\/} {\bf 48} 1365--1366

\bibitem{Yu2016}
Yu S, Li L, Shi G, Zhu C, Zhou X and Shi Y 2016 {\em Applied Physics Letters\/}
  {\bf 108} \urlprefix\url{https://doi.org/10.1063/1.4944789}

\bibitem{Zheng2016}
Zheng S, Zhang W, Zhang Z, Jin X, Chi H and Zhang X 2016 {\em Photonic
  Research\/} {\bf 4} B9--B13
  \urlprefix\url{http://www.osapublishing.org/prj/abstract.cfm?URI=prj-4-5-B9}

\bibitem{Thide2007}
Thid\'e B, Then H, Sj\"oholm J, Palmer K, Bergman J, Carozzi T~D, Istomin Y~N,
  Ibragimov N~H and Khamitova R 2007 {\em Phys. Rev. Lett.\/} {\bf 99}(8)
  087701 \urlprefix\url{https://link.aps.org/doi/10.1103/PhysRevLett.99.087701}

\bibitem{Collin1969}
Collin R~E and Zucker F~J 1969 {\em Antenna Theory\/} (McGraw-Hill Book
  Company, Inc., New York)

\bibitem{Orfanidis2016}
Orfanidis S~J 2016 {\em Electromagnetic Waves and Antennas\/}
  \urlprefix\url{http://www.ece.rutgers.edu/~orfanidi/ewa}

\bibitem{Chandran2004}
Chandran S 2004 {\em Adaptive Antenna Arrays\/} (Springer-Verlag, Berlin)

\bibitem{Haupt2010}
Haupt R~L 2010 {\em Antenna Arrays\/} (John Wiley and Sons, Inc., Hoboken, New
  Jersey)

\bibitem{Rabinovich2013}
Rabinovich V and Alexandrov N 2013 {\em Antenna Arrays and Automotive
  Applications\/} (Springer Science, Business Media, New York)

\bibitem{Abramowitz1972}
Abramowitz M and Stegun I~A 1972 {\em Handbook of Mathematical Functions\/}
  10th ed (Washington, DC: National Bureau of Standards, US Government Printing
  Office)

\bibitem{VolkeSepulveda2002}
Volke-Sepulveda K, Garc\'{e}s-Ch\'{a}vez V, Ch\'{a}vez-Cerda S, Arlt J and
  Dholakia K 2002 {\em Journal of Optics B: Quantum and Semiclassical Optics\/}
  {\bf 4} S82 \urlprefix\url{http://stacks.iop.org/1464-4266/4/i=2/a=373}

\bibitem{Knudsen1956}
Knudsen H 1956 {\em IRE Transactions on Antennas and Propagation\/} {\bf 4}
  452--472

\bibitem{Zhu2013}
Zhu J, Cai X, Chen Y and Yu S 2013 {\em Optics Letters\/} {\bf 38} 1343--1345
  \urlprefix\url{http://ol.osa.org/abstract.cfm?URI=ol-38-8-1343}

\bibitem{Zhu2014}
Zhu J, Chen Y, Zhang Y, Cai X and Yu S 2014 {\em Optics Letters\/} {\bf 39}
  4435--4438 \urlprefix\url{http://ol.osa.org/abstract.cfm?URI=ol-39-15-4435}

\bibitem{Born1999}
Born M and Wolf E 1999 {\em Principles of {Optics}\/} 7th ed (Cambridge
  University Press)

\bibitem{Miller2000}
Miller D~A~B 2000 {\em Applied Optics\/} {\bf 39} 1681--1699
  \urlprefix\url{http://ao.osa.org/abstract.cfm?URI=ao-39-11-1681}

\bibitem{Edfors2012}
Edfors O and Johansson A~J 2012 {\em IEEE Transactions on Antennas and
  Propagation\/} {\bf 60} 1126--1131

\bibitem{Devaney2012}
Devaney A~J 2012 {\em Mathematical Foundations of Imaging, Tomography and
  Wavefield Inversion\/} (Cambridge University Press)

\bibitem{Jackson1999}
Jackson J~D 1999 {\em Classical Electrodynamics\/} 3rd ed (John Wiley and Sons,
  Inc., New York)

\bibitem{Morse1953}
Morse P~M and Feshbach H 1953 {\em Methods of Theoretical Physics\/}
  (McGraw-Hill Book Company, Inc., New York)

\bibitem{Golub1996}
Golub G~H and Van~Loan C~F 1996 {\em Matrix Computations\/} 3rd ed (The Johns
  Hopkins University Press, Baltimore)

\bibitem{Horn1994}
Horn R~A and Johnson C~R 1994 {\em Topics in Matrix Analysis\/} 2nd ed
  (Cambridge University Press)

\bibitem{Harrington2001}
Harrington R~F 2001 {\em Time-harmonic electromagnetic fields\/} (John Wiley
  and Sons, Inc., New Jork)

\end{thebibliography}

\providecommand{\newblock}{}

\end{document}